\documentclass[12pt,preprint]{aastex}
\renewcommand{\cite}{\citealp}

\newcommand{\rrl}{{RR~Lyrae}}

\shorttitle{Variable stars in Fornax 5}
\shortauthors{Greco et al.}

\begin{document}

\title{Variable Stars in the Fornax dSph Galaxy. III. The Globular Cluster Fornax 5\altaffilmark{1}}

\author{
Claudia Greco,\altaffilmark{2,3} 
Gisella Clementini,\altaffilmark{2}
M\'arcio Catelan,\altaffilmark{4,5,6}
Enrico V. Held,\altaffilmark{7}
Ennio Poretti,\altaffilmark{8}
Marco Gullieuszik,\altaffilmark{7} 
Marcella Maio,\altaffilmark{2}
Armin Rest,\altaffilmark{9}
Nathan De Lee,\altaffilmark{10}
Horace A. Smith,\altaffilmark{11}
Barton J. Pritzl\altaffilmark{12}
}

\altaffiltext{1}{This paper includes data gathered with the 6.5 m Magellan telescope located at 
Las Campanas Observatory, Chile, 
and WFPC2 at HST archival data.}

\altaffiltext{2}{INAF, Osservatorio Astronomico di
Bologna, via Ranzani 1, I-40127 Bologna, Italy;
gisella.clementini@oabo.inaf.it}

\altaffiltext{3}{Current address: Observatoire de Geneve, 51, ch. Des Maillettes, CH-1290 Sauverny, Switzerland; claudia.greco@unige.ch}

\altaffiltext{4}{Departamento de Astronom\'ia y Astrof\'isica, 
Pontificia Universidad Cat\'olica de Chile, Av. Vicu\~na Mackenna 4860,
782-0436 Macul, Santiago, Chile; mcatelan@astro.puc.cl}

\altaffiltext{5}{John Simon Guggenheim Memorial Foundation Fellow} 

\altaffiltext{6}{On sabbatical leave at Michigan State University, 
Department of Physics and Astronomy, 3215 Biomedical and Physical 
Sciences Bldg., East Lansing, MI 48824} 

\altaffiltext{7}{INAF, Osservatorio Astronomico di
Padova, vicolo dell'Osservatorio 5, I-35122 Padova, Italy;
enrico.held@oapd.inaf.it, marco.gullieuszik@oapd.inaf.it}

\altaffiltext{8}{INAF, Osservatorio Astronomico di
Brera, via E. Bianchi 46, 23807 Merate, Italy;
ennio.poretti@brera.inaf.it}


\altaffiltext{9}{Physics Department, Harvard University, 17 Oxford 
Street, Cambridge, MA 02138; arest@physics.harvard.edu}

\altaffiltext{10}{Department of Astronomy, University of Florida, 211 Bryant Space Center, Gainesville, FL 31611-2055;
ndelee@astro.ufl.edu}

\altaffiltext{11}{Department of Physics and Astronomy, Michigan State University, East Lansing, MI 48824-2320;
smith@pa.msu.edu}

\altaffiltext{12}{University of Wisconsin Oshkosh, Physics and Astronomy Department, Oshkosh, WI 54901;
pritzlb@uwosh.edu}


\begin{abstract}
We present a new study of the variable star population
in globular cluster 5 of the
Fornax dwarf spheroidal galaxy, based on 
$B$ and $V$ time series photometry obtained with 
the MagIC camera of the 6.5 m Magellan Clay telescope 
and complementary 
HST archive data. 
Light curves and 
accurate
periodicities were obtained 
for 30 RR Lyrae stars and 1 SX Phoenicis variable. 
The RR Lyrae sample includes 15 fundamental-mode pulsators, 13 first-overtone pulsators, 1 candidate double-mode pulsator
and one RR Lyrae star with uncertain type classification.

The average and minimum periods of the ab-type RR Lyrae stars, 
{\rm $\langle Pab\rangle$}= 0.590 days, $P_{ab,min}$=0.53297 days
and the position
in the horizontal branch type--metallicity plane, indicate that the cluster has Oosterhoff-intermediate 
properties, 
basically confirming previous indications by 
Mackey \& Gilmore (2003b), although with some differences both in the period and type 
classification of individual variables.

The  average apparent magnitude of the Fornax~5 RR Lyrae stars is  
 {\rm $\langle V{\rm (RR)}\rangle$}=21.35 $\pm$0.02 mag 
($\sigma$=0.07 mag, average on 14 stars more likely belonging to the cluster, and having well sampled
light curves). 
This value leads to a true distance modulus of $\mu_0$=20.76$\pm$0.07 
($d=141.9^{+4.6}_{-4.5}$ kpc)  
if we adopt for the cluster the metal abundance by Buonanno et al. 
(1998; [Fe/H]=$-2.20 \pm 0.20$),
or $\mu_{0}$=20.66$\pm$0.07 ($d=135.5^{+4.4}_{-4.3}$ kpc),  
if we adopt Strader et al.'s (2003) metal abundance  
([Fe/H]=$-1.73 \pm 0.13$).

\end{abstract}

\keywords{
galaxies: dwarf
---galaxies: individual (Fornax)
---globular clusters: individual (Fornax~5)
---stars: horizontal branch 
---stars: variables: other 
---techniques: photometry 
}

\section{Introduction}
This is the third in a series of papers devoted to the study
of the variable star population of the Fornax dwarf spheroidal galaxy (dSph)
based on a photometric survey that reaches $V \sim $ 25-26 mag and covers 
about 1~deg$^2$ of the galaxy field and all the 5 globular clusters belonging to Fornax 
(see Clementini et al. 2006, for a description of the whole project). 
In Greco et al. (2007, Paper I) we have presented 
the first 
study of the variable star population of Fornax~4 (hereafter For~4), the cluster located towards the centre of Fornax, which
Hardy (2002) and Strader et al. (2003) suggest might be the galaxy nucleus.
In Poretti et al. (2008, Paper II) 
we reported results on the identification of 85 SX Phoenicis stars 
in a $33^{\prime} \times 34^{\prime}$ North 
portion of the Fornax dSph field. In the present paper we present results for variable stars 
in the globular cluster Fornax~5 (hereafter For~5), that, along with Fornax 1 (hereafter, For~1), is 
the most external of the Fornax clusters. In the following papers of the series we will report on the study of the brighter variables 
(RR Lyrae stars and Anomalous Cepheids)  
in the Fornax field, 
as well as in the globular clusters Fornax~3 (hereafter, For~3, Clementini et al. 2009, in preparation), For~1 and  
Fornax~2 (hereafter, For~2).

The 5 globular clusters of the Fornax dSph galaxy  
(Hodge 1961, 1965, 1969) form a complex system with
no straightforward correlations existing between galactocentric distance,
metallicity and age. They are all metal poor
($-2.0 \le$ [Fe/H] $\le -1.5$), have rather similar mass and are distributed at varying distances from the 
 galaxy center marked by For~4. 
For~5 is placed at the galaxy periphery, 
about 26$^{\prime}$ North and 
33.5$^{\prime}$ East of For~4.
According to the surface brightness profile 
it seems a post-collapse cluster with core radius 
$r_c = 2.08 \pm 0.17$ arcsec (Mackey \& Gilmore 2003a). 
The literature values for the cluster tidal radius range 
from $r_t= 64 \pm 18$ arcsec (Webbink 1985), to $r_t= 74 \pm 5$ arcsec 
(Demers et al. 1994), to $r_t= 91$ arcsec (Smith et al. 1996).

Buonanno et al. (1998) published a color magnitude diagram 
(CMD) for the cluster based on {\it Hubble Space Telescope} ({\it HST}) Wide Field Planetary Camera 2 (WFPC2; Program ID 5917)  
 observations, from which they estimated the cluster relative age and metallicity. 
The CMD is rich in stars,
with a horizontal branch (HB) extending 
to the blue across the RR Lyrae star instability strip, 
 and suggests that For~5 
is coeval with the oldest Galactic globular clusters. 
 Buonanno et al. (1998) also find that For~5 is metal poor, with [Fe/H]=$-2.20 \pm 0.20$ dex, 
 on the Zinn \& West (1984) metallicity scale. 
 This metallicity is in good agreement with previous estimates, from both photometric (Buonanno et al. 1985)  
and spectroscopic studies (Dubath, Meylan \& Mayor 1992).

Strader et al. (2003) published new age and metal abundance estimates for the cluster 
based on low-resolution, integrated Keck spectra. They  
derive a larger metallicity of [Fe/H]=$-1.73 \pm 0.13$ (on the Zinn \& West 1984 metallicity scale). 
They also suggest that For~5 is a few billion years younger than the other clusters in Fornax. 

Buonanno et al. (1998) found evidence for 
36 possible variables in For~5, on the basis of the star frame-to-frame variations in their
HST data, but did not provide identification or measurement of any variables.
More recently, 
Mackey \& Gilmore (2003b; hereafter MG03) presented a more quantitative  
study of the 
RR Lyrae stars in For~5,  
based on the same HST dataset used by Buonanno et al. (1998) to build the cluster CMD.
The small number (14-16 images) and short time interval (8-9 consecutive hours) spanned  
by the HST archive data were inadequate to allow MG03 a direct determination of the periods, and they fitted template light 
curves of RR Lyrae stars to the data to get an estimate of periodicity and amplitude of the light variation.  
Based on their period estimates MG03 concluded that the RR Lyrae stars in For~5 have mean 
pulsation period intermediate between the two Oosterhoff types (Oosterhoff 1939)
observed within the MW GCs. 

In this paper we present 
results from a new study of the variable star population in For~5, that 
resulted in the identification and in the definition of accurate 
pulsation parameters for 30 \rrl\ stars and one SX Phoenicis (SX Phe) variable. 
Seven of the RR Lyrae stars and the SX Phe variable are new discoveries, that do not have 
a counterpart in the MG03 study.
 The periods, amplitudes, and period-amplitude distributions of the  \rrl\ stars allowed us to firmly 
establish the
Oosterhoff type (Oo-type) of the cluster. 
Their average luminosity was used to measure the cluster distance.

Observations and data reductions are presented in Sect.~2. 
The variable star identification is described 
in Sect.~3. The period search technique, the pulsation properties of the For~5 variable stars,  
and their membership in the parent cluster are discussed in Sect.~4. In this section we also compare our results with the
MG03 study. 
In Sect.~5 we discuss the Oosterhoff classification 
of For~5 and compare it to For~4, for which in Paper I we found an Oosterhoff-Intermediate behavior. 
 In Sect.~6 we derive the cluster distance from the average luminosity of the
 RR Lyrae stars. Final results are summarized in Sect.~7. 

\section{Observations and data reductions}

$B,V$ time series photometry of For~5\  was obtained  over two consecutive nights in November 2003 with the MagIC camera of the 
Magellan/Clay 6.5-m telescope, equipped with a 2048 $\times$ 2048 SITe CCD having resolution of 
0.069 arcsec pixel$^{-1}$. Observations of For~4 were 
also obtained
during the same nights (see Paper I).
We covered a total field of view (FOV) of 2.4$^{\prime}$ $\times$ 2.4$^{\prime}$,
 centered 
on the cluster at
 R.A.=2$^h$ 42$^m$ 21.07$^s$, DEC=$-$34$^{\circ}$ 06$^{\prime}$ 07.5$^{\prime \prime}$ \ (J2000). 
 According to the tidal radii provided by Webbink (1985) and Demers et al. (1994) for For~5, our
FOV should cover the cluster almost in its entirety. 
Nights were photometric with seeing conditions varying from  0.45$\arcsec$ to  $0.65 \arcsec$.
We acquired 52 $V$ and 20 $B$ images of For~5.   
 Average exposure times were of 
500 seconds for the $V$ frames and 700 s for the $B$ ones.
Observations of the standard fields T Phe and Ru 149 (Landolt 1992) 
were obtained to calibrate the 
data to the standard Johnson-Cousins photometric system.

The Magellan frames were bias-subtracted and flatfield-corrected using the MagIC tool in 
IRAF\footnote{IRAF is distributed by the National Optical Astronomical 
Observatories, which are operated by the Association of Universities for
Research in Astronomy, Inc., under cooperative agreement with the 
National Science Foundation}. PSF photometry was performed with DAOPHOT-ALLSTAR and 
time series photometry for all the stars was produced with ALLFRAME (Stetson 1994). 
The internal photometric precision at the HB level is of 0.01 mag.
The photometric calibration of the For~5 data (as well as for For~4, Paper I) 
was derived using the standard stars
observed at the Magellan telescope on the night of Nov. 13, 2003 (UT).  
A set of linear calibration relations was derived according to the procedure described in Paper I,
to which the interested reader is referred to for details.
The zero point uncertainties of the calibration relations are estimated
of the order 0.05 mag in $B$ and $V$, and 0.03 mag in $(B-V)$.

The Magellan time series observations of 
For~5\ correspond to a total observing time of 26000 sec in $V$ and 
 14000 sec in $B$ and
 produced a CMD reaching $V \sim 26 $ mag.
 This CMD is shown in Fig.~1. 
No reliable photometry could be obtained for stars inside a 
radius of about $8 \arcsec$ from the cluster center due to crowding. 
The CMD in Fig.~1 was obtained cutting off this central region. 
	    
In order to improve the light curve sampling and the period definition, 
the Magellan data were complemented by 
the HST archival data of HST Program 5917, the same dataset used in the Buonanno et al. 
(1998) and in the MG03 studies.    
The HST archival dataset 
 comprises 14 F555W and 16 F814W images obtained over about 8-9 consecutive hours, in 1996, June, 
 with the planetary camera (PC; resolution 0.0455 arcsec~pixel$^{-1}$, FOV 36$^{\prime}$ $\times$ 36$^{\prime}$) and
 the 3 wide field (WF) cameras (resolution 0.0996 arcsec~pixel$^{-1}$, FOV 80$^{\prime}$ $\times$ 80$^{\prime}$)
 of the WFPC2. 
Exposure times varied from
 160 to 600 sec for the F555W filter, and from 20 to 900 sec for the F814W filter. 
 The reduction of the HST archival data was performed with DAOPHOT-ALLSTAR-ALLFRAME exactly in the same 
 way as for the Magellan dataset. The F555W WFPC2 instrumental magnitudes were linked to the Johnson-Cousins magnitudes 
 of the Magellan dataset following the procedure described at the beginning of Section 4. 
 Logs of the observations and details of the instrumental set-ups are provided in
Table~1 for the Magellan and HST data, separately. 

\section{Variable star identification}

Variable stars were identified by applying the Image Subtraction Technique of ISIS2.1 (Alard 2000)
 to the Magellan $V$ and $B$ dataset  separately.
 Images in each photometric band were first aligned and re-mapped onto the same grid. Then 
 $V$ and $B$ reference images were constructed by stacking the frames obtained in best seeing conditions, and 
each individual frame of the time-series was subtracted from the corresponding reference image,  
after convolution with a suitable kernel to match seeing variations and geometrical distortions on the
individual images. The output of ISIS2.1 is a median image (``var.fits") of all difference frames, in which non-variable
objects disappear and candidate variable stars stand out as bright/dark peaks.
The last step of ISIS2.1 is to perform profile fitting photometry of the variable sources in the subtracted frames to
build differential flux light curves for each candidate variable star identified in the var.fits frame
(see Alard 2000, for details). 
To exclude false detections, the $B$ and $V$ catalogues of sources showing light variations according to ISIS were cross-identified 
in order to select objects
which consistently varied in both photometric bands.
About 300 candidate variables were identified in the field of For~5, including the cluster central region. 
The final selection of bona-fide variables was made through visual inspection of 
the light curves in differential flux. A very large fraction of the
candidate variables turned out to be 
spurious detections due to CCD bad columns, hot pixels, satellite trails, and 
saturated objects. The spurious origin of their variability was revealed by the 
almost identical light curves and from the close positions on the frames of the contamined 
objects. 
The ISIS catalog was cross-identified against the DAOPHOT/ALLFRAME photometric catalog obtaining 
light curves in magnitude scale from the ALLFRAME time series photometry and confirming the variability for 21 stars.
A further 10 variable stars were recovered with ISIS in the central region of For~5, where we lack reliable 
{\sc psf} photometry of the Magellan data. These stars have only differential flux light curves,  
in the Magellan dataset.
By matching the coordinates the confirmed variables identified on the Magellan images
were cross-identified against the photometry of the HST archival data. 
 Twenty-six out of the 31 variable stars found in For~5 have a counterpart in the HST photometry.
Of the missing five ones, four (namely stars V22, V26, V29 and V30) are outside the FOV covered by the HST observations, and one 
star (V5) lacks a secure counteridentification in the HST dataset since it lies in a very crowded region 
close to For~5 center (see also Section 4.3).

\section{Period Search and Properties of the Variable Stars}

\subsection{Analysis of the light curves}

The time series of the candidate variable stars were analyzed  
using GrATiS (Graphical Analyzer of Time Series), a custom 
software developed at the Bologna Observatory by P. Montegriffo (see Di Fabrizio 1999, Clementini et al. 2000) which uses 
both the Lomb periodogram (Lomb 1976, Scargle 1982) and the best fit of the data with
a truncated Fourier series (Barning 1963). 
Both  the Magellan and the HST datasets  
were used, when available, to determine the pulsation characteristics of the variable stars 
 from the study of the light curves. In order to combine the HST F555W and the Magellan $v$ photometries 
 we used a number of stable reference stars common to both datasets and having colors close to those
 of the candidate variables, and built instrumental differential light curves. These were then tied to the 
 Magellan instrumental photometric system (by adding up the instrumental magnitude of the comparison stars in the
 Magellan photometry) and finally calibrated  to the standard system using the calibration procedure discussed in Sect.~2.
The time series $B,V$ photometry of the confirmed variable stars with light curves
in magnitude scale is provided in Table~2.
The full version of this
table 
is  published  in the electronic edition of the Journal. 
Identification and coordinates of the confirmed variable stars are presented 
in Table~3.

 We were able to derive reliable periods for the vast majority of the confirmed variable stars detected in 
 For~5. However, the accuracy of our
 period determinations
 varies significantly (in the range from 2 to 6 digits) depending on the spectral window
 of the available data, whether only Magellan or 
 Magellan + HST (see Table~3).   

 The Magellan observations of For~5 consist of 3.5 and 7.6 hours of consecutive 
 exposures taken along the two nights of 2003, November 13 and 14 (UT), respectively
from HJD=2452956.538 to HJD=2452956.684, and from HJD=2452957.518 to HJD= 2452957.836, 
while 
the HST data were acquired along about 7.8 consecutive hours in 1996, June 4-5 (UT)
(from HJD=2450238.721 to HJD=2450239.048).
Magellan as well as HST photometry is available for 24 of the 31 variables we found in For~5. 
They include 9 of the variables having light curves only in differential flux in the 
Magellan dataset.  
 By comparing the Magellan differential flux light curves and the HST light curves in 
 magnitude scale we were able
to improve the period determination for these variables, which, however, remains more uncertain for
stars with periods longer than 0.5 days.
Moreover, using the HST photometry we managed 
to obtain a rough estimate of their  $\langle V \rangle$ magnitudes that confirmed
they are HB 
variables, we estimated the $V$ amplitudes of 4 of them, and could set lower limits to the amplitude of other 4 variables. 

The definition of the period of the For~5 variables is 
driven by the observations taken on the two consecutive nights at the Magellan telescope. 
The For 5 field was surveyed for 0.46~days on these nights,
allowing us to pin down periods shorter than this value in
a very reliable way, whilst longer periods are more uncertain.
The HST data improved the phase coverage in several cases,
also refining the values of the Magellan periods.
The spectral window of the For~5 data shows several peaks
at multiple integers of $\pm$0.00037~d$^{-1}$ around the central
peak, owing to the large gap between the HST and the
Magellan observations. Their power is higher than 98\% 
as far as  $\pm$0.05~d$^{-1}$ from the central peak. 
Up to six-digits periods were  necessary to fold the light curves
of some of the variables listed in Table~3, but the alias structure
introduces larger uncertainties. For instance, the data of V13
can be fit in equivalent manner by periods of
0.399790, 0.406060 or 0.411842~days and we cannot select the true period in an unambiguous way. 
 However, given the small difference between alternative periods the scientific discussion 
is not significantly affected in this case. Similarly, among the long--period RR Lyrae stars, the light
curve of V17 is fitted by $P=0.68$ days but  
we cannot completely rule out a longer period for this star, while 
the data of V20 are fitted equally well by periods of 0.74740
and 0.769393~days. In the latter case, the period difference is not negligible and we took the period uncertainties into account 
in the scientific discussion. The uncertain periods have been flagged in Table~3. 
In a few cases where a shorter period corresponding to  
first-overtone pulsation, and a longer period corresponding to fundamental mode pulsation, were both likely, we used  
the shape, amplitude and time-rise to the maximum of the light curve (if available) to achieve the final classification, 
which generally  turned out to be as a
fundamental mode pulsator. 

The cluster variable stars include 30 RR Lyrae stars and one SX Phe variable.
The RR Lyrae sample comprises 15 {\it ab-}, 13 {\it c-}type, 1 candidate 
double-mode pulsator ({\it d-}type, star V18), and a further RR Lyrae star, variable V15, 
whose period definition and type classification are uncertain. 
The properties (pulsation period, type, time of maximum light, number of phase points in the
$V$ and $B$ bands, intensity-averaged ${\rm \langle V\rangle}$ and
${\rm \langle B\rangle}$ magnitudes, and amplitudes of the light variation, A$_V$ and A$_B$) 
of the variable stars are summarized in Table~3.
Examples of light curves are shown in Fig.~2. The complete atlas of the RR Lyrae 
stars light curves is presented in Figs. 3 and 4. 
The light variation of the For~5 RR Lyrae stars is generally well sampled in both the
$V$ and the $B$ photometric bands, with standard deviations of the least--square fits of the light curves 
in the range from 0.02 to 0.05 mag, and from 0.02 to 0.06 mag in $V$ and $B$, respectively.
  
The scatter and, specifically, the two maxima with slightly different shape observed 
in the light curves of V18 on two consecutive nights suggest that 
the star could be a double-mode \rrl\ variable. 
Indeed, we find that a solution with fundamental mode 
period of $P_0$= 0.49602 days and period ratio 
of $P_1$/$P_0$=0.74640, is possible, although not as the best solution. In fact, 
the phase coverage is incomplete and the peak of one term is perturbed by
the alias structure of the other. 
 It is also noteworthy that
there are other RRc variables in For~5 with longer periods than V18 that do not seem
to be double-mode pulsators, V19 with a slightly longer period than V18
shows a little scatter and some evidence for a double maximum light, but V28, V26 and V13, with significantly longer periods,
seem to show little sign of an excess light curve scatter. 
Additional data are needed to establish  the double mode nature of V18, that, if confirmed, 
would bring to 7 the number of RRd stars detected so far in extragalactic globular clusters:  4 are reported in the
Reticulum cluster of the Large Magellanic Cloud (Ripepi et al. 2004) and two RRd's were found in Fornax 4
by Greco et al. (2007, Paper I). The frequency of RRd stars in the Fornax dSph clusters and what the presence 
of RRd stars in 
extragalactic clusters may 
imply will be discussed in more detail in a following paper of this series (Clementini et al. 2009, in preparation).

V29 is the only short--period ($P <$ 0.1 days) variable detected in our survey of For~5. 
Despite the star's faintness ($V\sim$23.7),  
the peak at $f$=20.20~d$^{-1}$ stands out clearly in the amplitude
spectrum of the $V$ data, with a  signal-to-noise ratio of 7.0. 
Being located beyond 45 arcsec from the cluster center, the star 
probably belongs to the Fornax field. 
The period (0.0495~days) and the $\langle  B \rangle$ and $\langle V \rangle$ magnitudes (23.97 and
23.74 mag, respectively) put the star close to the $P-L$ relation  
of the Fornax SX Phe stars  
(see Fig.~7 in Paper~II; V29 is similar to 2\_V3796).

\subsection{Variable stars in the CMD}

The variable stars are plotted in the CMD of Fig.~1 according to their 
magnitudes and colors intensity-averaged over the full pulsation cycle, and with 
different symbols/colors for the first-overtone, fundamental, double-mode RR Lyrae stars
and the SX Phe variable. Thanks to the good sampling of the light curves the RR Lyrae 
instability strip in For~5 is very well defined and appears to not extend much in color 
(see Fig.~1). The blue edge of the strip is at 
${\rm \langle B\rangle} - {\rm \langle V\rangle}$=0.22$\pm$0.03 mag, as derived from the 
average of the three bluest {\it c-}type RR Lyrae stars with secure membership to the
cluster (see Sect.~4.4), namely, V21, V23 and V24.
The instability strip red edge appears to be at 
${\rm \langle B\rangle} - {\rm \langle V\rangle} \sim $ 0.40 mag, 
with a rather sharp separation
between RR Lyrae and red HB stars.  
The color of the blue edge of the RR Lyrae strip can be used to estimate the
reddening of the cluster. By matching the blue edge of the RR Lyrae 
strip in M3 [$E(B-V)_{\rm M3}$=0.01 mag; Harris 1996] to the colors of the bluest RR Lyrae stars
in For~5  we derive: $E(B-V)$=0.04$\pm 0.03$ mag.
This  value can be compared with the value of 
$E(B-V)$=0.022 mag for the foreground Galactic extinction in the direction of 
 For~5, derived from the Schlegel, Finkbeiner, \& Davis (1998) maps.  

\subsection{Comparison with previous work}

MG03  have identified 40 RR Lyrae in For~5, of which 33 are on the PC, 2 on the WFPC2-2, 1 on the WFPC2-3, 
and 4 on the WFPC2-4. We have identified 30 RR Lyrae and one SX Phe star in For~5, of which 23 have a counterpart in MG03,
namely 19 stars on the PC, 2 on the WFPC2-2, one star on the WFPC2-3, and one star on the WFPC2-4.
 The cross-identification of the For~5 variables with MG03 is provided
in Table~4. The table comprises 22 variables classified as RR Lyrae stars in both studies and one star (V19) listed  
among MG03 candidate variables (see Table~3 of MG03) which we classify as a first-overtone RR Lyrae.
Four out of the 22 RR Lyrae stars in common with MG03 are 
classified {\it ab-}type pulsators by these authors, but have very uncertain periods. 
We confirm the type classification for all of them, but derive longer periods 
by about $\Delta P \sim$0.16 days, on average, with full range from 0.03 to 0.23 days. 
Further three stars classified as
first-overtones by MG03 are instead fundamental-mode pulsators (V10 and V11) and a candidate
double-mode pulsator (V18). We were not able to unambiguously establish the period and type
classification of V15. MG03 classify this star as an RRc with P=0.394 days. We find that P=0.340
phases our data (see Fig. 4); however, other periods for the star are also possible. 
For the remaining 14 stars in common there is agreement in the type
classification, 5 are {\it ab-}type and 9 are {\it c-}type 
pulsators according to both studies. Among the {\it ab-}type pulsators, three 
have periods underestimated by 0.07 days, on average, with full range from 0.01 to 0.14 days; 
while 2 have periods overestimated by 0.05 days in MG03.
Among the {\it c-}type pulsators, for V31 we fully confirm MG03's period determination, four have periods underestimated by 0.02 days, on
average, with full range from 0.01 to 0.05 days; 
while 4 have periods overestimated by 0.005 days, on average, with full 
range from 0.001 to 0.01 days, in MG03. The comparison between our period determinations and MG03 values for the stars
in common is shown in Fig. 5, where we plot the difference between our and MG03 estimates as a function of 
our periodicities. Open and filled circles represent first-overtone and fundamental-mode pulsators, respectively,
according to our type classification.
In our sample there are 3 variables on the PC (namely, stars V1, V5, V17) and one variable star on the
WFPC2-4 (V24) that have no counterpart in MG03. 
V1 and V5 are in the very central region of For~5. They were detected with the image 
subtraction technique, however, no reliable measure of magnitude could be obtained for them 
either in the HST or in the Magellan dataset. Star V17 has very small amplitudes.
These reasons may explain why MG03 missed these three variable stars.   
Finally an additional 4 variables in our sample (namely, stars V22, V26, V29, and V30) are outside the field covered 
by the HST observations.
 
\subsection{Membership Probability}

We checked the contamination of the For~5 RR Lyrae stars by field variables. 
Given the different stellar content of the globular cluster and
the surrounding Fornax field populations, and the presence of a stellar
population gradient in Fornax, estimating the expected number of cluster
RR\,Lyrae variables from the surface density profile in our small
observed field is not straightforward.
To illustrate the difference between the stellar populations of the
cluster and the field, we have selected stars in two regions, inner and
outer. Then we have used the outermost variable stars in the 
field to empirically
estimate the expected surface density of RR\,Lyrae stars in the field of the Fornax galaxy.

The count excess produced by the stars in the globular cluster For~5
is very small  beyond $R=45\arcsec$. This is consistent with the
density profile of For~5 obtained by Mackey \& Gilmore (2003a).  
 In our Magellan data, 2 RR Lyrae stars (one fundamental-mode, V30, and 
 one first-overtone, V31, the 
 bluest RR Lyrae in the field of For~5, with 
${\rm \langle B\rangle} - {\rm \langle V\rangle}$=0.15 mag) have been identified beyond
$45\arcsec$, in an area of $1.4 \times 10^4$  arcsec$^2$.  On the basis
of the overall surface density profile of For~5, for statistical purposes we assume that these
variables are not associated with the cluster.  This provides an {\it
upper limit} to the number density of field variable stars.
 Then we estimate a surface density of $1.5 \times 10^{-4}$ RR Lyrae per 
arcsec$^2$ in the adjacent Fornax field. 
In the inner region, within a radius of $45\arcsec$ from the center,
0.9 field RR Lyrae stars are therefore expected.  

We note that the limiting radius 45 arcsec adopted for the membership discussion
is significantly smaller than the cluster tidal radius. In fact, 
many stars outside this arbitrary chosen limiting radius may well individually belong 
to the cluster.

\section{Oosterhoff type and metallicity}

The pulsation properties of the For~5 RR Lyrae stars are summarized in Table~5 where we 
list the total number of variable stars with well established type 
classification (${\rm N{tot}}$),
the  number of fundamental mode (${\rm N{ab}}$), first overtone (${\rm N{c}}$) 
and double mode pulsators (${\rm N{d}}$),   
the ratio of number of RRc to total
number of RR Lyrae stars,  
the average periods of the {\it ab-} and 
{\it c-}type RR Lyrae stars,  
the shortest and longest RRab periods,  
and the shortest and longest RRc periods. 

The average period of 
the For~5 {\it c-}type RR Lyrae stars is:
{\rm $\langle Pc\rangle$}=0.356 days ($\sigma$=0.041 days, 
average on 11 stars and excluding the double mode pulsator) or 
{\rm $\langle Pc\rangle$}=0.358 days ($\sigma$=0.040 days, 
average on 12 stars) if the  double mode pulsator is included.
The average period of the {\it ab-}type RR Lyrae stars is  
{\rm $\langle Pab\rangle$}=0.590 days ($\sigma$=0.039 days, average on 
 13 stars with certain period determination).
 If we include in the mean also star V20, whose period is 
 uncertain, with two
alternative values  $P$=0.7474 and 0.769393 days equally possible,
the average period becomes: {\rm $\langle Pab\rangle$}=0.601 days ($\sigma$=0.056 days, average on 14
stars) if we adopt for V20 the P=0.7474 days period, or  
{\rm $\langle Pab\rangle$}=0.602 days ($\sigma$=0.061 days, average on 14
stars) if we adopt for V20 the P=0.769393 days period.  
The shortest period RRab is star V14 with $P_{ab,min}$=0.53297 days, and the
longest period RRab is star V20. 
Here, we have considered only variables within 45 $\arcsec$ from the center of For~5, 
since they are more likely cluster members (see Sect.~4.4).  
According to the average period of the fundamental-mode RR Lyrae stars For~5, like For~4 
(see Paper I),  appears to be an
Oosterhoff-intermediate (Oo-Int)
cluster, an Oo-type that has no counterpart among the MW GCs\footnote{We recall that in the MW,
GCs that contain significant numbers of RR Lyrae stars divide into two distinct groups according to the
average periods of the fundamental-mode and first-overtone pulsators, and the fraction of first-overtone
over total number of RR Lyrae stars (${\rm f_c}$). Namely, OoI clusters have   
{\rm $\langle Pab\rangle$}=0.55, {\rm $\langle Pc\rangle$}=0.32, ${\rm f_c}$=0.17, while OoII clusters
have {\rm $\langle Pab\rangle$}=0.64, {\rm $\langle Pc\rangle$}=0.37, 
and ${\rm f_c}$=0.44 (Oosterhoff 1939, Clement et al. 2001).}, 
but rather common instead among
the extragalactic clusters (Catelan 2004, 2009). We thus confirm, with improved reliability,
the cluster Oosterhoff classification suggested by MG03.

The period-amplitude distributions of the For~5 RR Lyrae stars  
is shown in Fig.~6, where expanded symbols are used for stars located at distances less than 
 45$\arcsec$ from the center of For~5.  
Figure~6 shows that the majority of the  {\it ab-}type RR Lyrae stars in  For~5
appears to be on the OoI line; however, a number of RRab's  are found near the OoII location.
It is also noteworthy that 
the average period of the first-overtone pulsators is more similar
to those of the OoII Galactic GCs. Furthermore, 
if we restrict ourselves to variables with more
secure membership, 
the ratio of first-overtone pulsators (RRc) over total
number of RR Lyrae stars (${\rm f_c}$ = 0.41/0.44) 
is closer to the typical value of an OoII cluster.

The metallicity of For 5 is somewhat 
uncertain, being [Fe/H]=$-2.20 \pm 0.20$ dex for Buonanno et al. (1998), and
[Fe/H]=$-1.73 \pm 0.13$ dex for Strader et al. (2003), on the Zinn \& West (1984) metallicity scale, and a 
systematic difference of 0.47 dex between the two independent studies.
 To get some hint of the cluster metallicity we have fit the  
CMD of For~5 to the mean ridgelines of the Galactic globular clusters
M15 (NGC7078; [Fe/H]$_{\rm M15}=-2.15 \pm 0.08$, Zinn \& West 1984), 
and M3 (NGC5272; [Fe/H]$_{\rm M3}=-1.66  \pm 0.06$, Zinn \& West 1984), taken from Durrell \& Harris (1993) 
for M15 
 and Ferraro et al. (1997) for M3, shifted in magnitude and color 
to match the
For~5 CMD main branches. 
This comparison is shown in Fig.~1, where the red solid line is the ridgeline of M15 and the
blue dashed line is the ridgeline of M3. 
To match the For~5 CMD we have shifted the M15 ridgeline by $\Delta V = +5.50$~mag in magnitude, and  
$\Delta (B-V)=-0.05$~mag in color. Since the reddening of M15 is $E(B-V)_{\rm M15}=0.10 \pm 0.01$ mag (Durrell \& Harris 1993),
the color shift required to match For~5 to M15 implies for the cluster a 
reddening $E(B-V)_{\rm For~5}=0.05 \pm 0.01$ mag, in good agreement with the value derived 
from the color of the blue edge of the RR Lyrae instability strip (see Sect. 4.2). 
Similarly, M3 ($E(B-V)_{\rm M3}=0.01 \pm 0.01$ mag, Harris 1996) was shifted by 
$\Delta (B-V)=+0.04$~mag in color to match the reddening of For~5, and by  $\Delta V = +5.70$~mag in magnitude.
%
The CMD of For~5 is reproduced by the ridgeline of M15, while the M3 ridgeline 
is about 0.1 mag too red than that of For~5.  This comparison suggests that the metal abundance of For~5 is closer to the metallicity
of M15 and lower than that of M3, in agreement with the conclusions of Buonanno et al. (1998).

Metallicity estimates can also be derived from the cluster RR Lyrae stars, by using a variety of methods either 
based on the shape of the light curve of individual stars, or on the mean pulsation properties
of the sample. 
We have used the parameters of the Fourier decomposition of the $V$-band light curves, along with
Jurcsik \& Kov\'acs (1996) method for the RRab stars (equation~3 in their paper),  and 
Morgan, Wahl \& Wieckhorst (2007) formula for the RRc stars (equation~3 in their paper), to estimate
individual metallicities, with accuracy of 
about 0.3 dex, for the For~5 RR Lyrae stars which have well sampled and regular light curves. 
Results are summarized in Table~6. The corresponding average metal abundance, for stars within 45$^{\prime \prime}$ from 
the cluster center, is:
${\langle {\rm [Fe/H]\rangle}} = - 1.97$ dex ($\sigma$=0.16, average on 9 stars), on the 
Zinn \& West (1984) metallicity scale. 
Using instead relations from Sandage (2006), we find [Fe/H]=$-$2.10 dex from the mean 
period of the For~5 fundamental mode RR Lyrae stars ({\rm $\langle Pab\rangle$}=0.590 days);
 and $-$2.33 dex from the shortest 
period RRab, V14, 
a star located within $45\arcsec$ from the cluster center. 
Finally, following Sandage (1993), we find [Fe/H]=$-$1.86/$-1.88$ dex, from the mean period of the 
RRc/RRd stars  
within $45\arcsec$ from the cluster center. All these determinations are on the Zinn\& West (1984) metallicity scale.
Thus, different methods based on the cluster RR Lyrae stars 
seem to suggest a low
metal abundance for For~5, closer to that of Buonanno et al. (1998). 

Nevertheless, we note that none of the conclusions on the Oo-Int status of For~5
are affected in any way by a change in the cluster metal abundance from $-2.2$ to $-1.7$~dex since,
independently of the adopted metal abundance, the {\rm $\langle Pab\rangle$} value places
For~5 in the Oo-Int region of the
$\langle P_{ab}\rangle - {\rm [Fe/H]}$ diagram (Pritzl et al. 2002;
Catelan 2009). 
In particular, if ${\rm [Fe/H]} = -1.7$ dex as derived by Strader et al. (2003), For~5 would fall  
right in the middle of the distribution of the Oo-Int clusters, 
whereas adopting ${\rm [Fe/H]}=-2.2$ dex actually places the cluster slightly 
to the right of this distribution (see Figs. 6 and 9 of Catelan 2009). In fact, 
and as discussed in Catelan et al. (2009, in preparation), the key 
quantities
defining Oosterhoff status appear to be the average and minimum
   periods of the {\it ab-}type RR Lyrae stars. In terms of these two 
   quantities For~5 appears entirely consistent with an
   Oo-Int classification. 

We have also reconsidered the HB morphology of For~5 by computing the Lee-Zinn parameter 
(Zinn 1986; Lee 1990; Lee, Demarque \& Zinn 1990), $(B\!-\!R)/(B\!+\!V\!+R)$
(where $B$,
$R$, and $V$ represent the numbers of blue, red, and variable stars
on the HB, respectively) of For~5, based on our photometry and number counts. 
We have considered only stars with 8$^{\prime \prime} < r < 45^{\prime \prime}$,
since we do not have reliable photometry of the Magellan data within
8$^{\prime \prime}$. Our results are: $R$=8$\pm 1$, $V$=16$\pm 1$,  
$B$=33$\pm 1$, and $(B\!-\!R)/(B\!+\!V\!+R)$=0.44 $\pm 0.17$ in agreement with
Buonanno et al. (1998) who derived 0.44$\pm$0.09, and slightly smaller than the MG03 values, who find 0.52$\pm$0.04
(all chips) and 0.52$\pm$0.05 (PC only).
According to our $(B\!-\!R)/(B\!+\!V\!+R)$ value 
For~5 locates on the right edge of the distribution of the Oo-Int clusters 
band in the [Fe/H]-HB type 
diagram (see Fig.~8 in Catelan 2009), 
for [Fe/H]=$-2.20$ dex, and on the middle
of this band for [Fe/H]=$-1.73$ dex.
We note that the MG03 values would instead place For~5 slightly outside the
Oo-Int band for [Fe/H]=$-2.20$ dex (see Fig. 8 of Catelan 2009). 

The properties of the RR Lyrae star populations in the Fornax dSph GCs 
are summarized
in Table~7. 
Results for For~5 from the present work, and for For~4 
from Paper I, are reported in the upper portion of the table. They correspond 
to variable stars whose membership to the parent clusters is more certain  
and contamination by field stars is minimized. 
Results for For~5 from MG03 are reported instead in the lower part 
of the table for comparison with the present 
study. In the last three rows of the table we also summarize the MG03 results for 
For~1, For~2 and For~3. We will revise and update the numbers for these three clusters 
in the following papers of this series. 
There is generally good agreement between our results for For~5 and those of MG03,
when for the latter, we restrict to variable stars which MG03 consider to have good period 
estimates. Some differences are found instead among the 
parameters of the fundamental-mode RR Lyrae stars derived in the two studies, since, 
as discussed
in Sect.~4.3, MG03 generally tend to underestimate the period of the 
{\it ab-}type RR Lyrae stars. MG03~~ ${\rm f_c}$ value based on all 
the candidate RR Lyrae stars differs largely from the  
value derived using only variables which MG03 consider to have good periods; however,  
it is interesting  that our ${\rm f_c}$ values are closer to the value based on the full sample of candidate  RR Lyrae stars that 
MG03 found in the cluster.
 
The globular clusters of the Fornax dSph appear all to have an Oosterhoff intermediate type,
 based on the mean period of their {\it ab-}type RR Lyrae stars, ${\langle P_{ab}\rangle}$
 (see Col.~3 of Table 7).
Indeed, they are {\it confined within} the Oosterhoff `gap' and span it almost entirely from 
its short period edge (with For~2) to its upper period edge 
(with For~3).  
These 5 globulars form a fairly homogeneous sample of Oosterhoff intermediate clusters and may  
contribute to unveiling the nature of the Oosterhoff intermediate type, namely whether it is just a 
  transition class or whether it is a distinct group with physical  properties common to all 
  Oo-Int clusters (see Catelan et al. 2009). 
However, the results for the Fornax clusters are not straightforward.
  We note that the mean periods of the  
{\rm c-}type RR Lyrae stars in these clusters are more similar to those of OoII Galactic 
clusters, and are generally longer than those  
of the Oosterhoff intermediate globular clusters in  the LMC.  
 On the other hand, the full range of $f_c$ values from OoI to OoII types is spanned by these clusters, thus 
 suggesting a weak 
 reliability of this parameter to discriminate among Oosterhoff types. 
The period-amplitude diagrams of the Fornax dSph clusters 
do not provide clearcut indications either, since in this plane 
the Fornax clusters show a variety of positions ranging from Oo~I (e.g. For~2, and For~4) to
Oo-Int/Oo II types, 
 thus suggesting that the
 Oosterhoff intermediate status is not necessarily accompanied by a clearcut intermediate (e.g., between
  OoI and Oo I types) position of the RR Lyrae stars in the Bailey diagram.
  Rather, the Fornax clusters seem to support the idea that the position of the blue edge for 
  first overtone pulsation ($P_{\rm ab,min}$) plays a
  dominant role in defining the Oosterhoff intermediate status.
  We will come back to this issue in Catelan et al. (2009).

\section{Cluster Distances} 

The average magnitude of the RR Lyrae stars in For~5 is 
$\langle V{\rm (RR)}\rangle$=21.35$\pm$0.02 mag ($\sigma$=0.07 mag, average on 14 stars
with well sampled light curves and more secure membership to the cluster).
This average value
changes only marginally if we also include in the sample 7 RR Lyrae stars in the central region 
of For~5 which only have HST photometry (see upper portion of Table~3) becoming 
$\langle V{\rm (RR)}\rangle$=21.34$\pm$0.02 mag ($\sigma$=0.09 mag, average on 21 stars).
Our {\rm $\langle V{\rm (RR)}\rangle$} values agree within their uncertainties with the Buonanno et al. (1998)
apparent magnitude of the cluster HB, $V_{\rm HB}{\rm (For~5)}=21.30 \pm 0.05$,  and with
the average magnitude of the 
full sample of candidate RR Lyrae stars (40 stars) detected in
For~5 by MG03, {\rm $\langle V{\rm (RR)}\rangle$}=21.33 $\pm 0.01$. 
 
In order to derive the distance to For~5 from the average luminosity 
of the RR Lyrae stars we need estimates of the cluster reddening
and metallicity, as well as values for the slope and zero point  
of the  RR Lyrae luminosity-metallicity relation. 
For consistency with the analysis of For~4 in Paper I we adopt an absolute magnitude of $M_V$ = 
0.59$\pm$0.03 mag for RR Lyrae stars of [Fe/H] =$-$1.5 dex (Cacciari \& Clementini, 2003),
and $\Delta M_V/{\rm [ Fe/H]}$=0.214$\pm$0.047 mag/dex (Clementini et al. 2003, Gratton et al.
2004)
 for the slope of the luminosity-metallicity relation.  
We then adopt a standard extinction law (A$_V$=3.1$\times E(B-V)$) 
and a reddening value  
of 0.05 $\pm$ 0.01 mag, which was  
derived as the weighted average of four independent estimates, namely: (i) the
$E(B-V)$ value of MG03, $E(B-V)$=0.03$\pm 0.01$ (intrinsic) $\pm$0.02 (systematic) mag; 
(ii) the Buonanno et al.(1998) $E(V-I)$ transformed to $E(B-V)$,  
$E(B-V)$=0.06$\pm 0.05$ mag; (iii) the reddening we have obtained by matching the 
cluster CMD to the mean ridgelines
of M15, $E(B-V)$=0.05 $\pm$ 0.01 mag (see Sect. 5); and (iv) the reddening
we derive by matching the blue edge of the RR Lyrae strip in M3
to the colors of the bluest RR Lyrae stars
in the cluster, $E(B-V)$=0.04$\pm 0.03$ mag (see Sect. 4.2). 
The distance modulus of For~5 is then 
$\mu_{0}$(For~5)=20.76$\pm$0.07 
($d=141.9^{+4.6}_{-4.5}$ kpc) if we adopt for the cluster the metal abundance by Buonanno et al. 
(1998; [Fe/H]=$-2.20 \pm 0.20$),
or $\mu_{0}$(For~5)=20.66$\pm$0.07 ($d=135.5^{+4.4}_{-4.3}$ kpc) 
if we adopt 
the Strader et al. (2003) metallicity ([Fe/H]=$-1.73 \pm 0.13$).
Errors in the cluster distance moduli are the sum in 
quadrature of uncertainties of 0.02 in {\rm $\langle V{\rm (RR)}\rangle$}
(dispersion of the average),
0.05 mag in the zero point of the photometry, 0.01 mag in $E(B-V)$ (corresponding to 0.03 mag in
$A_V$), and of 0.03 mag and 0.047 mag/dex, respectively, in the zero point and in the slope of the 
RR Lyrae luminosity-metallicity relation.
The cluster distance modulus does not change significantly if we use a different
RR Lyrae luminosity-metallicity relation, e.g. that of Chaboyer (1999) or that of 
Catelan \& Cort\'{e}s (2008).

Distances to the Fornax dSph GCs from the present study and from Paper I for 
For~4 are summarized in Table~8 and compared with distances from the MG03 study.
Our distance moduli for For~5 compare very well with 
the MG03 estimates particularly
if the low metallicity value is adopted for the cluster.  
 
 \section[]{Summary and conclusions}

We have obtained light curves and periods
for 30  RR Lyrae stars  (13 {\it ab-}, 14 {\it c-} and 1 {\it d-}type pulsator)
and 1 SX Phe variable in the globular cluster For~5 of the
Fornax dSph galaxy. 
According to our estimate of the surface density of RR Lyrae stars in the 
field of the Fornax dSph surrounding For~5, at least 28$\pm$1 
out of 30 RR Lyrae variables 
are cluster members.
The average periods of {\it ab-} and {\it c-}type RR Lyrae stars, 
and the minimum period of the {\it ab-}type pulsators 
point to an Oosterhoff-intermediate status for the cluster,
unlike what is seen for the vast majority of the Galactic globular clusters.
The previous
Oosterhoff-intermediate classification suggested by 
MG03  is confirmed with improved reliability. 
Our photometry and 
number counts also suggest a value of the Lee-Zinn HB type of
$(B\!-\!R)/(B\!+\!V\!+R)$ = 0.44$\pm 0.17$ for For~5.
This value places the cluster on the Oo-Int clusters band in the
[Fe/H]-HBtype diagram.

The average magnitude of the RR Lyrae stars that are more likely members of  For~5 is 
$\langle V{\rm (RR)}\rangle$= 21.35$\pm$0.02 mag, and leads to $\mu_0$=20.76 $\pm$ 0.07 mag
or $\mu_0$=20.66 $\pm$ 0.07 mag 
whether Buonanno et al. [Fe/H]=$-$2.20 dex, or Strader et al.
[Fe/H]=$-$1.73 dex, are 
adopted for the cluster metallicity. The pulsation properties of the RR Lyrae
stars support a low metal abundance for For~5 closer to the
Buonanno et al.
value, and the distance modulus of $\mu_0$=20.76 $\pm$ 0.07 mag 
($d=141.9^{+4.6}_{-4.5}$ kpc).
This distance modulus for the cluster agrees well with the value found by MG03.

\acknowledgments 
We thank the anonymous referee for comments.
This research was funded 
by PRIN INAF 2006 (P.I.: G. Clementini). Support for M.C. is provided by Proyecto Basal 
   PFB-06/2007, by FONDAP Centro de Astrof\'{i}sica 15010003, by Proyecto 
   FONDECYT Regular \#1071002, and by a John Simon Guggenheim Memorial 
   Foundation Fellowship. HAS thanks the National Science Foundation 
   for support under grant AST 0607249.

\begin{table*}
      \caption[]{Instrumental Set-up and Log of the Observations}
	 \label{t:obs}
	 $$
	\begin{array}{lllccc}
	   \hline
	    \hline
	   \noalign{\smallskip}
{\rm ~~~~~Dates}	& {\rm ~~~~Telescope}  & {\rm ~~Instrument} & {\rm N_ B} &{\rm N_ V}&{\rm N_ I}\\
{\rm ~~~~~(UT)}  &		       &		    &	         &          &          \\
	    \hline
{\rm 1996, June~4-5}   & {\rm ~~HST}      & {\rm  ~~~WFPC2/PC} & $\nodata$& 14& 16~\\
                     &                  & {\rm  ~~~WFPC2/WF} & $\nodata$& 14& 16~\\
{\rm 2003, Nov. 13-15}& {\rm ~~Magellan/Clay}& {\rm  ~~~MagIC}&        20~& 52&$\nodata$\\
\hline
	 \end{array}
	 $$
\label{t:table1}

	    \end{table*}

\begin{table}
\caption{$V,B$ Photometry of the For~5 Variable Stars with Light Curves on a Magnitude Scale}
\vspace{0.5 cm} 
\begin{tabular}{cccc}
\hline
\hline
\multicolumn{4}{c}{For~5 - Star V19 - {\rm RRc}} \\
\hline
{\rm HJD} & {\rm V}  & {\rm HJD } & {\rm B}\\ 
{\rm ($-$2450238)} & {\rm (mag)}  & {\rm ($-$2452956) } & {\rm (mag)}  \\
\hline
 0.708973 &    21.51   &  0.546462 &  21.95  \\
 0.711751 &    21.55   &  0.551370 &  21.91  \\
 0.714529 &    21.52   &  0.579396 &  21.87  \\
 0.717307 &    21.54   &  0.584338 &  21.82  \\
 0.788839 &    21.53   &  0.608903 &  21.74  \\
 0.796246 &    21.57   &  0.613521 &  21.65  \\
 0.835948 &    21.45   &  0.618555 &  21.65  \\
 0.843587 &    21.32   &  0.643554 &  21.51  \\
 0.922873 &    21.34   &  0.691377 &  21.43  \\
 0.930281 &    21.31   &  1.555353 &  21.80  \\
\hline 

\end{tabular}

\label{t:table3}
\medskip
 Table~\ref{t:table3} is published in its entirety in the electronic edition of the {\it Astrophysical Journal}. 
 A portion is shown here
 for guidance regarding its form and content.
\end{table}

\begin{table}
\tiny
\caption[]{Identification and Properties of the For~5\ \ Variable Stars.}
\label{t:fornax5}
     $$
         \begin{array}{llcclllclclccc}
	    \hline
            \hline
           \noalign{\smallskip}
           {\rm Name} & {\rm Id}  & {\rm \alpha } & {\rm \delta} &  {\rm Type} &~~~{\rm P} & 
	    ~~~{\rm Epoch}  & {\rm \langle V\rangle}  & ~~{\rm N_V} & {\rm \langle B\rangle}  &~~{\rm N_B} &
	    {\rm A_V} & {\rm A_B} & {\rm Notes}\\
            & &{\rm (2000)}& {\rm (2000)}& & ~{\rm (days)}& ($-$2450000) &{\rm (mag)}&
  &{\rm (mag)} & &{\rm (mag)}& {\rm (mag)}&\\
            ~~(a)& & & & & & & &   & & & & &\\
          \noalign{\smallskip}
            \hline
            \noalign{\smallskip}
{\rm V1 } &   {\rm LC}117  & 2:42:21.25  & -34:06:07.60 & {\rm RRab} & 0.635   &  2956.587 &  $\nodata$  &  54+11&$\nodata$ & $\nodata$   & >0.19    & $\nodata$ & (b,c) \\
{\rm V2 } &   {\rm LC}77   & 2:42:21.01  & -34:06:10.56 & {\rm RRc } & 0.4085  &  2957.700 &  21.26	 &  55+14&$\nodata$ & $\nodata$   & 0.50     & $\nodata$ & (b) \\
{\rm V3 } &   {\rm LC}103  & 2:42:21.40  & -34:06:08.64 & {\rm RRc } & 0.2842  &  2956.623 &  21.33	 &  55+14&$\nodata$ & $\nodata$   & \ge 0.37 & $\nodata$ & (b) \\
{\rm V4 } &   {\rm LC}166  & 2:42:21.34  & -34:06:03.19 & {\rm RRc } & 0.3635  &  2957.797 &  21.21	 &  55+14&$\nodata$ & $\nodata$   & 0.56     & $\nodata$ & (b) \\
{\rm V5 } &   {\rm LC}172  & 2:42:21.20  & -34:06:02.17 & {\rm RRab} & 0.56    &  $\nodata$ &  $\nodata$  &  55   &$\nodata$ & $\nodata$   & $\nodata$& $\nodata$ & (b,d) \\
{\rm V6 } &   {\rm LC}175  & 2:42:21.02  & -34:06:01.68 & {\rm RRab} & 0.65    &  $\nodata$ &  $\nodata$  &  55+11&$\nodata$ & $\nodata$   & $\nodata$& $\nodata$ & (b,c) \\
{\rm V7 } &   {\rm LC}136  & 2:42:20.54  & -34:06:05.83 & {\rm RRab} & 0.605   &  $\nodata$ &  21.48      &  55+14&$\nodata$ & $\nodata$   & >0.37    & $\nodata$ & (b) \\
{\rm V8 } &   {\rm LC}149  & 2:42:21.58  & -34:06:05.01 & {\rm RRab} & 0.5620  &  2957.623 &  21.48	 &  55+14&$\nodata$ & $\nodata$   & >0.59    & $\nodata$ & (b) \\
{\rm V9 } &   {\rm LC}98   & 2:42:21.63  & -34:06:08.83 & {\rm RRab} & 0.6310  &  2956.593 &  21.23	 &  55+14&$\nodata$ & $\nodata$   & \ge 0.84 & $\nodata$ & (b) \\
{\rm V10} &   {\rm LC}118  & 2:42:21.68  & -34:06:07.49 & {\rm RRab} & 0.6041  &  2956.603 &  21.26	 &  51+14&$\nodata$ & $\nodata$   & >0.78    & $\nodata$ & (b) \\
          &                &             &              &            &         &            &             &       &         &              &          &           &     \\ 
{\rm V11} &   3583	   & 2:42:20.51  & -34:06:10.69 & {\rm RRab} & 0.547034&  2957.600 &  21.38:	 &  52+12&  21.67:    &   ~20 &   >0.76  &    >0.75& \\
{\rm V12} &   2577	   & 2:42:21.67  & -34:06:04.63 & {\rm RRab} & 0.533555&  ~238.145 &  21.27	 &  51+14&  21.52  &   ~20 &\ge 0.88  & \ge 1.06& \\
{\rm V13} &   2537	   & 2:42:21.72  & -34:06:07.58 & {\rm RRc } & 0.406060&  ~237.193 &  21.37	 &  50+12&  21.68  &   ~20 &	0.35  &     0.41& \\
{\rm V14} &   2482	   & 2:42:21.79  & -34:06:01.59 & {\rm RRab} & 0.53297 &  ~237.615:&  21.39:	 &  50+14&  21.59: &   ~18 &\ge 0.56  & \ge 0.94& \\
{\rm V15} &   3845	   & 2:42:20.21  & -34:06:05.88 & {\rm RR }  &$\nodata$&  $\nodata$ &  21.20	  &  47+14&  21.54  &	~17 &\ge 0.16  & \ge 0.25& (e)\\
{\rm V16} &   3631	   & 2:42:20.47  & -34:05:59.59 & {\rm RRab} & 0.605949&  ~238.796 &  21.29	  &  49+14&  21.63  &	~20 &	 1.06  &     1.22& \\
{\rm V17} &   2252	   & 2:42:22.05  & -34:06:07.61 & {\rm RRab} & 0.68::  & $\nodata$  &  21.36:  	  &  49+12&  21.75:    &	~20 &	>0.23  &    >0.32& (f)\\
{\rm V18} &   3274	   & 2:42:20.87  & -34:06:20.18 & {\rm RRc/d} & 0.37023 &  ~238.720 &  21.37	  &  52+14&  21.66  &	~19 &  	 0.45  &     0.58& (g)\\
{\rm V19} &   2818	   & 2:42:21.39  & -34:05:52.21 & {\rm RRc } & 0.372043&  2957.782 &  21.36	 &  52+14&  21.67  &   ~20 &	0.43  &     0.60& \\
{\rm V20} &   3936	   & 2:42:20.09  & -34:06:18.66 & {\rm RRab} & 0.7474: &  2957.675 &  21.45	 &  51+13&  21.85:    &   ~18 &	0.63  & \ge 0.74& \\
{\rm V21} &   2663	   & 2:42:21.58  & -34:05:50.29 & {\rm RRc } & 0.307246&  ~237.875 &  21.40	 &  52+14&  21.64  &   ~18 &	0.65  &     0.88& \\
{\rm V22} &   2706	   & 2:42:21.53  & -34:06:25.23 & {\rm RRab} & 0.61    &  2957.715 &  21.40	 &  50~~~&  21.76     &   ~19 &	0.69  &     0.84& (h) \\
{\rm V23} &   4097	   & 2:42:19.86  & -34:05:56.09 & {\rm RRc } & 0.333292&  ~237.850 &  21.37	 &  47+14&  21.61  &   ~20 &	0.61  &     0.70& \\
{\rm V24} &   1673	   & 2:42:22.84  & -34:06:11.51 & {\rm RRc } & 0.313763&  ~237.685 &  21.35	 &  52+13&  21.54  &   ~19 &	0.37  &     0.50& \\
{\rm V25} &   4350	   & 2:42:19.52  & -34:05:48.13 & {\rm RRc } & 0.365929&  2957.715 &  21.38	 &  50+14&  21.65  &   ~18 &	0.49  &     0.68& \\
{\rm V26} &   5057	   & 2:42:18.29  & -34:06:13.03 & {\rm RRc } & 0.389   &  2955.858 &  21.28	 &  50~~~&  21.54  &   ~19 &	0.48  &     0.73& (h)\\
{\rm V27} &   970	   & 2:42:24.04  & -34:06:11.83 & {\rm RRab} & 0.588055&  ~238.152 &  21.30:	 &  52+13&  21.65:  &	~20 &	 0.99: &    >0.63& \\
{\rm V28} &   4853	   & 2:42:18.69  & -34:05:43.30 & {\rm RRc } & 0.3768  &  2957.637 &  21.46	 &  51~~~&  21.76  &   ~18 &	0.30  &     0.34& (i)\\
{\rm V29} &   3427	   & 2:42:20.64  & -34:05:43.59 & {\rm SX~Phe}& 0.04950&  2957.755 &  23.74      &  42~~~&  23.97   &	~17 &    0.21  &     0.33 (h)& \\
{\rm V30} &   5472	   & 2:42:17.46  & -34:07:16.14 & {\rm RRab} & 0.538   &  2956.547 &  21.40	 &  52~~~&  21.68  &   ~20 &	1.19  &     1.48& (h)\\
{\rm V31} &   157	   & 2:42:25.64  & -34:05:07.30 & {\rm RRc } & 0.396   &  2957.844 &  21.21	 &  49+10&  21.36  &   ~19 &	0.55  &     0.67& \\
															   
\hline															   
            \end{array}
	    $$
{\tiny $^{\mathrm{a}}$ Variable stars were assigned increasing numbers starting from the cluster 
center, which was set at 
R.A.=2$^h$ 42$^m$ 21.07$^s$, DEC=$-$34$^{\circ}$ 06$^{\prime}$ 07.5$^{\prime \prime}$, (J2000).
Stars from V1 to V28 are located within 45$\arcsec$ from the For~5 center, hence are more likely cluster members
(see Sect.~4.4).}\\
{\tiny $^{\mathrm{b}}$ Light curves from the Magellan observations (50-55 data points) are available in differential flux only.
Magnitudes, when available 
are only from the HST dataset (10-14 data points). They were obtained through a rough calibration of the
HST F555W data done neglecting the color terms, they provide only a crude estimate of the actual average magnitude of these
variable stars, which often  have poor light curve coverage in the HST dataset.}\\
{\tiny $^{\mathrm{c}}$ The light curve coverage of V1 and V6 in the HST dataset is not sufficient to provide an estimate of the
star average $V$ magnitude.}\\
{\tiny $^{\mathrm{d}}$ V5 has no secure counterpart in the HST photometry.}\\
{\tiny $^{\mathrm{e}}$ Average magnitude and shape of the light curve are consistent with a small amplitude RR Lyrae star, but the period definition 
is very uncertain. A period of about 0.49 days phases 
the star light curve, a shorter period around 0.34 days is also possible. MG03 derive for the star a period of 0.394 but with
large uncertainty. In Fig.~4 the star light curves were folded according to the period P=0.34 which is closer to the value
found by MG03.}\\
{\tiny $^{\mathrm{f}}$ Shape and amplitude of the light curve indicate a long period fundamental mode RR Lyrae star, but the period definition 
is very uncertain. A period around 0.68 days phases 
the light curve well but with a large gap around the minimum light. Longer periods around 0.73 or 0.79 days are also possible.}\\
{\tiny $^{\mathrm{g}}$ The scatter and specifically the two maxima slightly different in shape observed 
in the light curves of V18 in two consecutive nights suggest that 
the star could be a double-mode \rrl\  variable.}\\ 
{\tiny $^{\mathrm{h}}$ These stars fall outside the FOV covered by the HST observations.}\\
{\tiny $^{\mathrm{i}}$ The HST photometry of V28 is systematically brighter than the Magellan photometry.}\\


\end{table}

\begin{table*}
\small
\caption[]{Cross-identification with MG03 study of the variable stars in For~5.}
\label{t:fornaxx5}
     $$
         \begin{array}{llllllll}
	    \hline
            \hline
           \noalign{\smallskip}
            {\rm Name} & {\rm Id}  & {\rm HST~Camera} &{\rm Type} &~~~{\rm P} & 
	    {\rm Id}  & {\rm Type} &~~~{\rm P} \\
	    {\rm This~paper}& {\rm This~paper}& &{\rm This~paper}& {\rm This~paper}& {\rm MG03} &
	     {\rm MG03} & {\rm MG03} \\  
	                    &                 &     &           &    {\rm (days)}       &            &
	                    &  {\rm (days)}       \\  
	    \noalign{\smallskip}
            \hline
            \noalign{\smallskip}
{\rm V2 } &   {\rm LC}77   &{\rm PC} &{\rm RRc } & 0.4085  & {\rm F5V15}  &{\rm RRc}  & 0.362	   \\
{\rm V3 } &   {\rm LC}103  &{\rm PC} &{\rm RRc } & 0.2842  & {\rm F5V03}  &{\rm RRc}  & 0.278	   \\
{\rm V4 } &   {\rm LC}166  &{\rm PC} &{\rm RRc } & 0.3635  & {\rm F5V11}  &{\rm RRc}	& 0.364      \\
{\rm V6 } &   {\rm LC}175  &{\rm PC} &{\rm RRab} & 0.65    & {\rm F5V06}  &{\rm RRab}	& 0.595      \\
{\rm V7 } &   {\rm LC}136  &{\rm PC} &{\rm RRab} & 0.605   & {\rm F5V38}  &{\rm RRab}	& 0.392^{a}  \\
{\rm V8 } &   {\rm LC}149  &{\rm PC} &{\rm RRab} & 0.5620  & {\rm F5V37}  &{\rm RRab}	& 0.532^{a}  \\
{\rm V9 } &   {\rm LC}98   &{\rm PC} &{\rm RRab} & 0.6310  & {\rm F5V25}  &{\rm RRab}	& 0.494      \\
{\rm V10} &   {\rm LC}118  &{\rm PC} &{\rm RRab} & 0.6041  & {\rm F5V10}  &{\rm RRc}	& 0.340      \\
{\rm V11} &   3583	   &{\rm PC} &{\rm RRab} & 0.547034& {\rm F5V22}  &{\rm RRc}	& 0.402      \\ 
{\rm V12} &   2577	   &{\rm PC} &{\rm RRab} & 0.533555& {\rm F5V16}  &{\rm RRab}	& 0.520      \\
{\rm V13} &   2537	   &{\rm PC} &{\rm RRc } & 0.406060& {\rm F5V29}  &{\rm RRc}	& 0.420      \\
{\rm V14} &   2482	   &{\rm PC} &{\rm RRab} & 0.53297 & {\rm F5V32}  &{\rm RRab}	& 0.603      \\
{\rm V15} &   3845	   &{\rm PC} &{\rm RR  } &$\nodata$& {\rm F5V05}  &{\rm RRc}	& 0.394      \\
{\rm V16} &   3631	   &{\rm PC} &{\rm RRab} & 0.605949& {\rm F5V26}  &{\rm RRab}	& 0.571      \\
{\rm V18} &   3274	   &{\rm PC} &{\rm RRc/d } & 0.37023 & {\rm F5V20}  &{\rm RRc}    & 0.422      \\
{\rm V19} &   2818	   &{\rm PC} &{\rm RRc } & 0.372043 & {\rm F5C02}^{b}  & $\nodata$   & $\nodata$     \\
{\rm V20} &   3936	   &{\rm PC} &{\rm RRab} & 0.7474: & {\rm F5V39}  &{\rm RRab}	& 0.513^{a}  \\
{\rm V21} &   2663	   &{\rm PC} &{\rm RRc } & 0.307246& {\rm F5V02}  &{\rm RRc}	& 0.311      \\
{\rm V23} &   4097	   &{\rm PC} &{\rm RRc } & 0.333292& {\rm F5V28}  &{\rm RRc}    & 0.326      \\
{\rm V25} &   4350	   &{\rm WFPC2-2} &{\rm RRc } & 0.365929& {\rm F5V18}  &{\rm RRc}     & 0.361	   \\
{\rm V27} &   970	   &{\rm WFPC2-4} &{\rm RRab} & 0.588055& {\rm F5V30}  &{\rm RRab}    & 0.433^{a}  \\													  
{\rm V28} &   4853	   &{\rm WFPC2-2} &{\rm RRc } & 0.3768  & {\rm F5V23}  &{\rm RRc}     & 0.379	   \\
{\rm V31} &   ~157	   &{\rm WFPC2-3} &{\rm RRc } & 0.396  & {\rm F5V12}  &{\rm RRc}      & 0.396	   \\
\hline															   
            \end{array}
	    $$
{\small $^{\mathrm{a}}$ Uncertain period determination (MG03).}\\
{\small $^{\mathrm{b}}$ Candidate variable according to MG03 Table~3.}\\

\end{table*}

\begin{table*}
\small
\caption[]{Average quantities for the For~5 RR Lyrae stars}
     $$
         \begin{array}{ccccclllccc}
	    \hline
            \hline
           \noalign{\smallskip}
           {\rm N{tot}} & {\rm N{ab}} & {\rm N{c}}  & {\rm N{d}} & {\rm N{c}/N{tot}} & {\rm \langle P{ab}\rangle} &  
	   {\rm \langle P{c}\rangle}& {\rm P_{ab,min}} & {\rm P_{ab,max}}& {\rm P_{c,min}}&{\rm P_{c,max}}\\
           {\rm (a)}         & {\rm (a)}         &      {\rm (a)}     &    {\rm (a)}        &        {\rm (a, b, c)}              &  {\rm (a, d)} &             
                    {\rm (a, e)}                  &                    &                  &          &                  \\
                       &             &              &            &                   &  {\rm (days)}              &             
           {\rm (days)}             &   {\rm (days)}   &  {\rm (days)}   &  {\rm (days)}   & {\rm (days)} \\
            \noalign{\smallskip}
            \hline
            \noalign{\smallskip}
	    
29(27)&16(15)&12(11)& 1(1)      & 0.41(0.41)& 0.590             & 0.356   & 0.53297& 0.7474:& 0.2842  & 0.4085 \\
      &      &	    &	        & 0.45(0.44)&		        & 0.358   &        & 	    &	      &        \\
\hline
            \end{array}
	    $$
{\small $^{\mathrm{a}}$ Numbers in parentheses and the ${\rm \langle P_{ab}\rangle}$, ${\rm \langle P_{c}\rangle}$ values correspond only to
RR Lyrae stars whose membership to the clusters is more certain (namely, RR Lyrae stars within $45\arcsec$ from the cluster center).}\\
{\small $^{\mathrm{b}}$  ${\rm N_{tot}}$ is calculated only from RR Lyrae stars with well established type classification
(29 out of the 30 RR Lyrae stars in For~5).}\\
{\small $^{\mathrm{c}}$ Ratio of RRc to total number of RR Lyrae stars with (2nd row values)
and without (1st row values) the candidate RRd star in For~5.}\\
{\small $^{\mathrm{d}}$ The ${\rm \langle P_{ab}\rangle}$ value of For~5 does not consider the two long period {\it ab-}type RR Lyrae stars 
V17 and V20 which have uncertain
periods. If these two stars are included ${\rm \langle P_{ab}\rangle}$ becomes 0.606/0.608 days, depending on the period
adopted for V20, whether 0.7474 or 0.769393 days.}\\
{\small  $^{\mathrm{e}}$ First overtone average period with (2nd row values) and without
(1st row values) the candidate double mode star.}\\
\end{table*}

\begin{table*}
\small
\caption[]{Metallicities from the Fourier decomposition of the light curves of the RR Lyrae stars within 45 arcsec from the cluster center.}
\label{t:fornaxx6}
     $$
         \begin{array}{lllcc}
	    \hline
            \hline
           \noalign{\smallskip}
            {\rm Name} & {\rm Type} &~~~{\rm P} & {\rm \phi_{31}}  & {\rm [Fe/H]_{\rm ZW84}}\\  
                       &            &{\rm(days)}  &                  &                        \\  
	    \noalign{\smallskip}
            \hline
            \noalign{\smallskip}
{\rm V13} &   {\rm RRc }  & 0.406060& 4.27   & $-1.84$  \\
{\rm V19} &   {\rm RRc }  & 0.372043& 2.77   & $-2.11$  \\
{\rm V21} &   {\rm RRc }  & 0.307246& 1.79   & $-2.00$  \\
{\rm V23} &   {\rm RRc }  & 0.333292& 2.17   & $-2.06$  \\
{\rm V24} &   {\rm RRc }  & 0.313763& 3.03   & $-1.66$  \\
{\rm V25} &   {\rm RRc }  & 0.365929& 3.28   & $-2.03$  \\
{\rm V26} &   {\rm RRc }  & 0.389   & 3.22   & $-2.07$  \\
{\rm V27} &   {\rm RRab}  & 0.588055& 1.69   & $-1.81:$ \\  												   
{\rm V28} &   {\rm RRc }  & 0.3768  & 2.77   & $-2.11$  \\
\hline			  	 											    
            \end{array}   
	    $$
{\small  The metal abundance of the fundamental mode RR Lyrae star was transformed to the Zinn \& West (1984) metallicity
scale by applying the transformation relation provided by equation~4 in Jurcsik (1995).}\\

\end{table*}


\begin{table*}
\tiny
\caption[]{Properties of the RR Lyrae stars in the globular clusters of the Fornax dSph\label{t:ammassi}}
     $$
         \begin{array}{lc cc cl rr cc cc}
	    \hline
            \hline
           \noalign{\smallskip}
{\rm Cluster} & {\rm (B\!-\!R)/(B\!+\!V\!+R)}&  {\rm \langle P{ab}\rangle}  &  {\rm \langle Pc\rangle}  &  {\rm P_{ab,min}}  &  {\rm P_{c,max}}  & 
 {\rm N{ab}}  &  {\rm Nc}  &  {\rm Nd}  &  {\rm f_c} &
{\rm Oo~Type}& {\rm Reference} \\
        &      {\rm (a)}          &     {\rm (b,days)}  & {\rm (days)}        &  {\rm (b,days)}            &   {\rm (b,days)}  &     {\rm (c)}  
        &       &         &  {\rm (d)}    &        & \\   
            \noalign{\smallskip}
            \hline
            \noalign{\smallskip}
{\rm For~5}  &  0.44       &0.590  	& 0.356/0.358  & 0.53297  &  0.409   &  15	  &  11   &       1 & 0.41/0.44	   &{\rm Oo-Int} & {\rm This~paper}\\
             &             &       	&	 &	    &	       &	  &	  &         &	           &		 &		   \\
{\rm For~4}  & -0.75       &0.594  	& 0.360  & 0.5191   &  0.400   &  16	  &  3    &  1      & 0.16	   &{\rm Oo-Int} & {\rm Paper~I}\\
             &             &       	&	 &	    &	       &	  &	  &         &    	   &		 &	       \\
{\rm For~5}  &	0.52(0.52) &0.577(0.532)& 0.374  & 0.494    &  0.445   &  8(13)   &  27   &$\nodata$& 0.77~(0.68)  &{\rm Oo-Int} & {\rm MG03}\\
             &             &            &  	 &	    & 	       &	  &	  &	 &	           &  	         &    	 \\
{\rm For~1}  & -0.30(-0.38)&0.611(0.546)& 0.431  & 0.570    &  0.459   &  5(10)   &  5    &$\nodata$& 0.50~(0.33)  &{\rm Oo-Int} & {\rm MG03}\\
{\rm For~2}  &  0.42(0.50) &0.574(0.494)& 0.373  & 0.512    &  0.404   &  15(30)  &  13   &$\nodata$& 0.46~(0.30)  &{\rm Oo-Int} & {\rm MG03}\\
{\rm For~3}  &	0.40(0.44) &0.613(0.532)& 0.404  & 0.479    &  0.510   &  32(61)  &  38   &$\nodata$& 0.54~(0.38)  &{\rm Oo-Int} & {\rm MG03}\\
	    
\hline
            \end{array}
	    $$
{\tiny $^{\mathrm{a}}$ {\rm $(B\!-\!R)/(B\!+\!V\!+R)$} values in parentheses correspond to MG03 counts for stars
on the PC only.}\\
{\tiny $^{\mathrm{b}}$  The ${\rm \langle P_{ab}\rangle}$ and P$_{ab,min}$  values from MG03 are those corresponding only to 
RRab stars with good periods according to these authors, the values in parentheses are from 
all the candidate RR Lyrae stars detected in MG03.} \\
{\tiny $^{\mathrm{c}}$ ${\rm N{ab}}$ values are the RRab's with good periods for MG03, values in parentheses correspond 
to the total number of candidate RR Lyrae stars in each cluster, according to the MG03 study.} \\
{\tiny $^{\mathrm{d}}$ f$_{c}$=Nc /(N${c}$ + N${ab}$), the f$_{c}$ values in parentheses correspond to the total number
of candidate RR Lyrae stars in each cluster, according to the MG03 study.}\\
\end{table*}

\begin{table*}
\tiny
\caption[]{Distance moduli for the globular clusters of the Fornax dSph galaxy based on
the average magnitude of the RR Lyrae stars\label{t:ammassi}}
     $$
         \begin{array}{lc cc ccl}
	    \hline
            \hline
           \noalign{\smallskip}
{\rm Cluster} & {\rm \langle V_{HB} \rangle}  &  {\rm E(B-V)} & {\rm [Fe/H]} & M_V {\it vs} [Fe/H]& \mu_0& {\rm Reference} \\
              &      {\rm (mag)}         &   {\rm (mag)}&             &     &  {\rm (mag)}      &          \\   
             \noalign{\smallskip}
            \hline
            \noalign{\smallskip}
{\rm For~5}  & 21.35\pm 0.02& 0.03\pm 0.01& -2.20\pm 0.20&0.214 \times {\rm ([Fe/H]+1.5)}+0.59&20.76\pm 0.07 &{\rm This~paper;~[Fe/H]~from~Buonanno~et~al.~(1998)} \\
             &              &             & -1.73\pm 0.13&                              &20.66\pm 0.07 &{\rm This~paper;~[Fe/H]~from~Strader~et~al.~(2003)}              \\
{\rm For~4}  & 21.43\pm 0.03& 0.10\pm 0.02& -2.01\pm 0.20&                              &20.64\pm 0.09 &{\rm Paper~I}     \\
             &              &             &              &	                        &              &      \\
{\rm For~5}  & 21.33\pm 0.01& 0.03\pm 0.01& -1.90\pm 0.06&0.23 \times {\rm ([Fe/H]+1.6)}+0.56&20.74\pm 0.05& {\rm MG03}\\
             &              &             &              &	                        &              &      \\
{\rm For~1}  & 21.27\pm 0.01& 0.07\pm 0.01& -2.05\pm 0.10&0.23 \times {\rm ([Fe/H]+1.6)}+0.56&20.58\pm 0.05& {\rm MG03}\\
{\rm For~2}  & 21.34\pm 0.01& 0.05\pm 0.01& -1.83\pm 0.07&                              &20.67\pm 0.05& {\rm MG03}\\
{\rm For~3}  & 21.24\pm 0.01& 0.04\pm 0.01& -2.04\pm 0.07&                              &20.66\pm 0.05& {\rm MG03}\\

\hline
            \end{array}
	    $$
\end{table*}

\clearpage

\begin{figure*} 
\includegraphics[width=16cm,bb=40 170 575 700,clip]{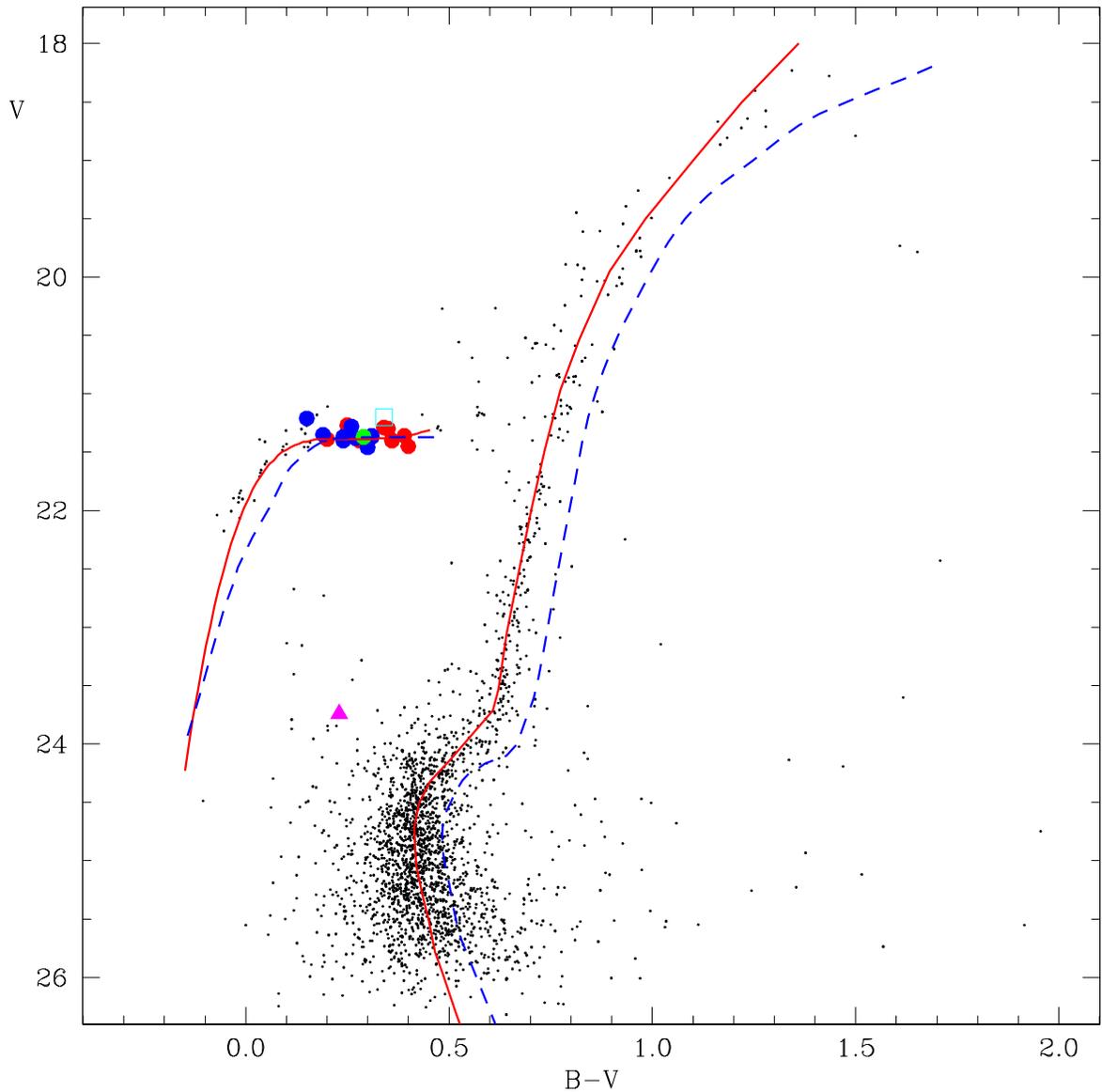}
\caption{Color-magnitude diagram of For~5\ from the Magellan data, showing stars located
outside an inner region of about 8$\arcsec$ from the cluster center. 
Variable stars are plotted according to their intensity-averaged magnitudes and colors, using  different colors/symbols
 for the different types.
{\it Red circles}: {ab-}type RR Lyrae stars; {\it blue circles}: first-overtone RR Lyrae stars; 
{\it green circle}: candidate double-mode pulsator (RRd); {\it square}: RR Lyrae star with uncertain type classification; 
{\it filled triangle}: SX Phe star. The bluest and brightest variable is V31, the 
most external of the For~5 RR Lyrae stars, which may belong to the 
field of the
Fornax dSph. The red (solid) line is the ridgeline of M15 from Durrell \& Harris (1993) and the blue (dashed) line is the ridgeline of M3
from Ferraro et al. (1997).}
\end{figure*}

\begin{figure}
\plotone{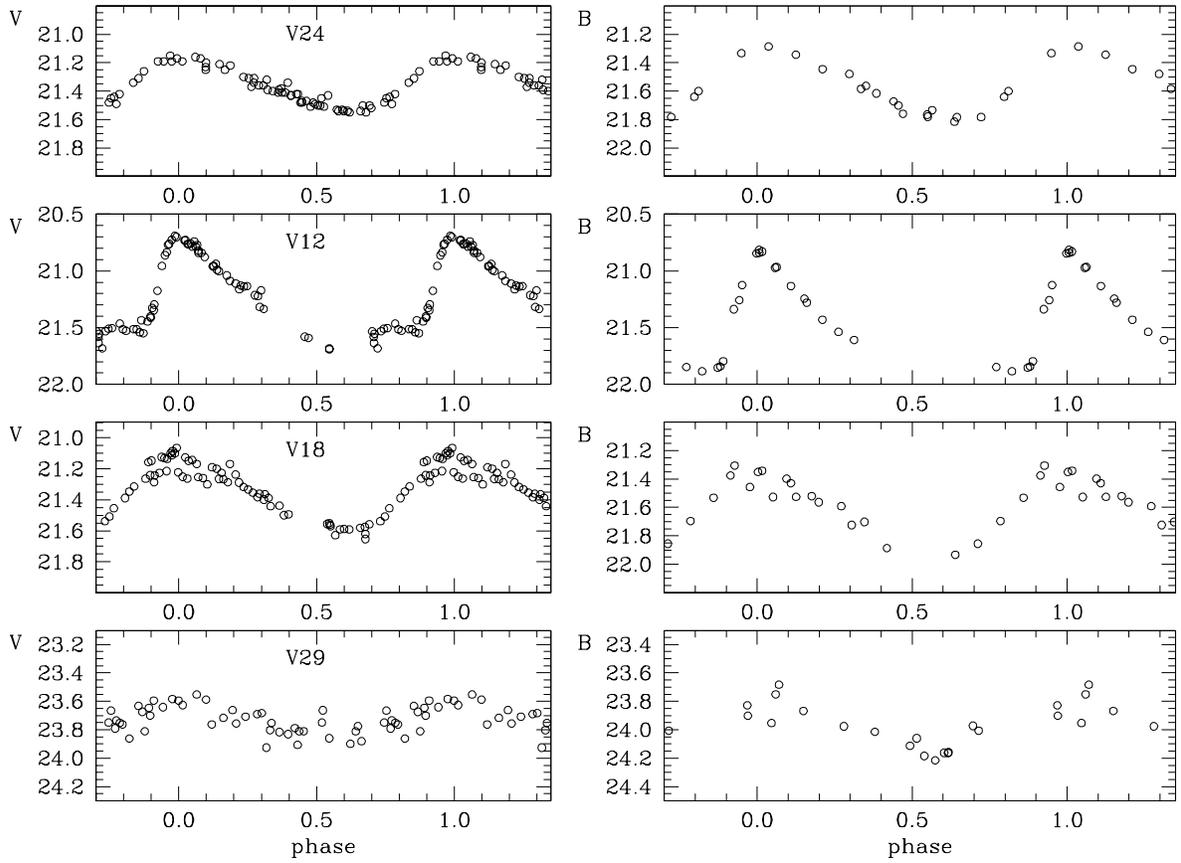}
\caption{$V$, $B$ light curves of variable stars in For~5. From top to bottom: {\it c-},  
{\it ab-}, {\it d-}type RR Lyrae stars, SX Phe star. 
}
\end{figure}

\begin{figure*} 
\includegraphics[width=18cm,bb=40 159 580 703,clip]{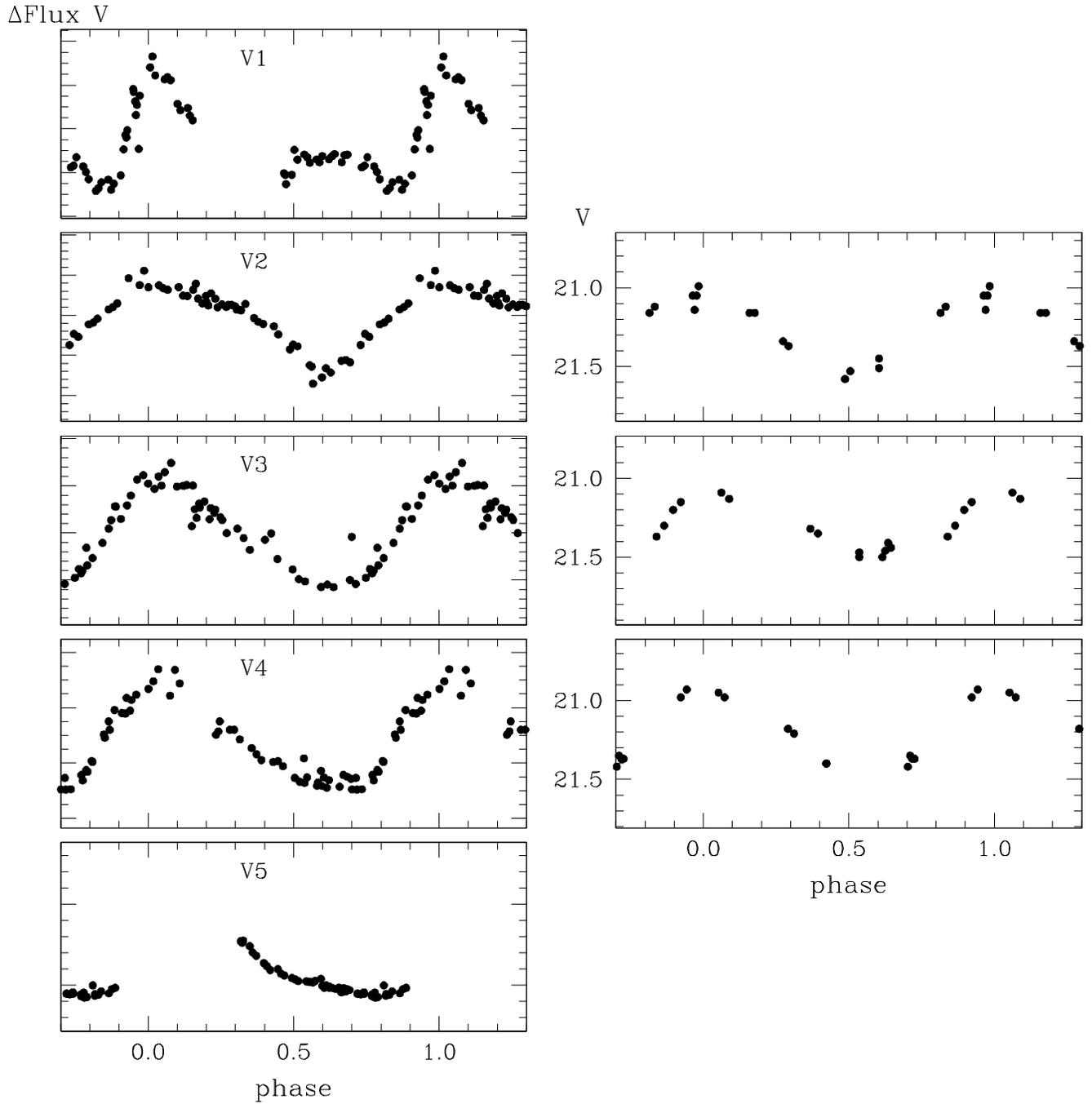}
\caption{Atlas of the light curves for \rrl\ stars  within 8$^{\prime \prime}$ from the 
center of For~5. {\it Left panel:} Light curves in $V$ differential flux, from the Magellan dataset.
{\it Right panel:} Light curves in $V$ magnitude, from the HST dataset.}
\end{figure*}

\begin{figure*} 
\figurenum{3}
\includegraphics[width=18cm,bb=40 159 580 703,clip]{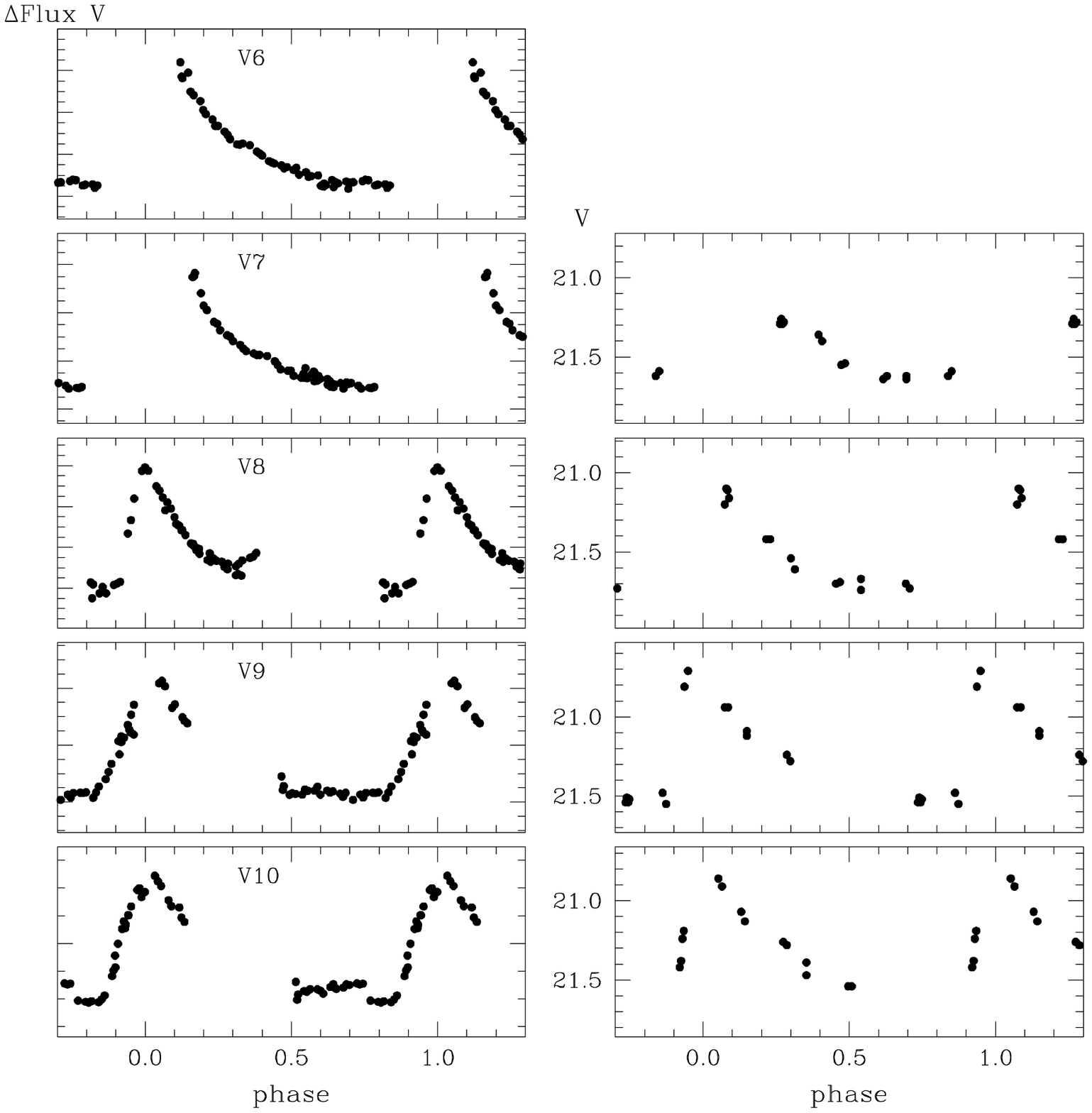}
\caption{continued}
\end{figure*}

\begin{figure*} 
\includegraphics[width=18cm,bb=40 159 580 703,clip]{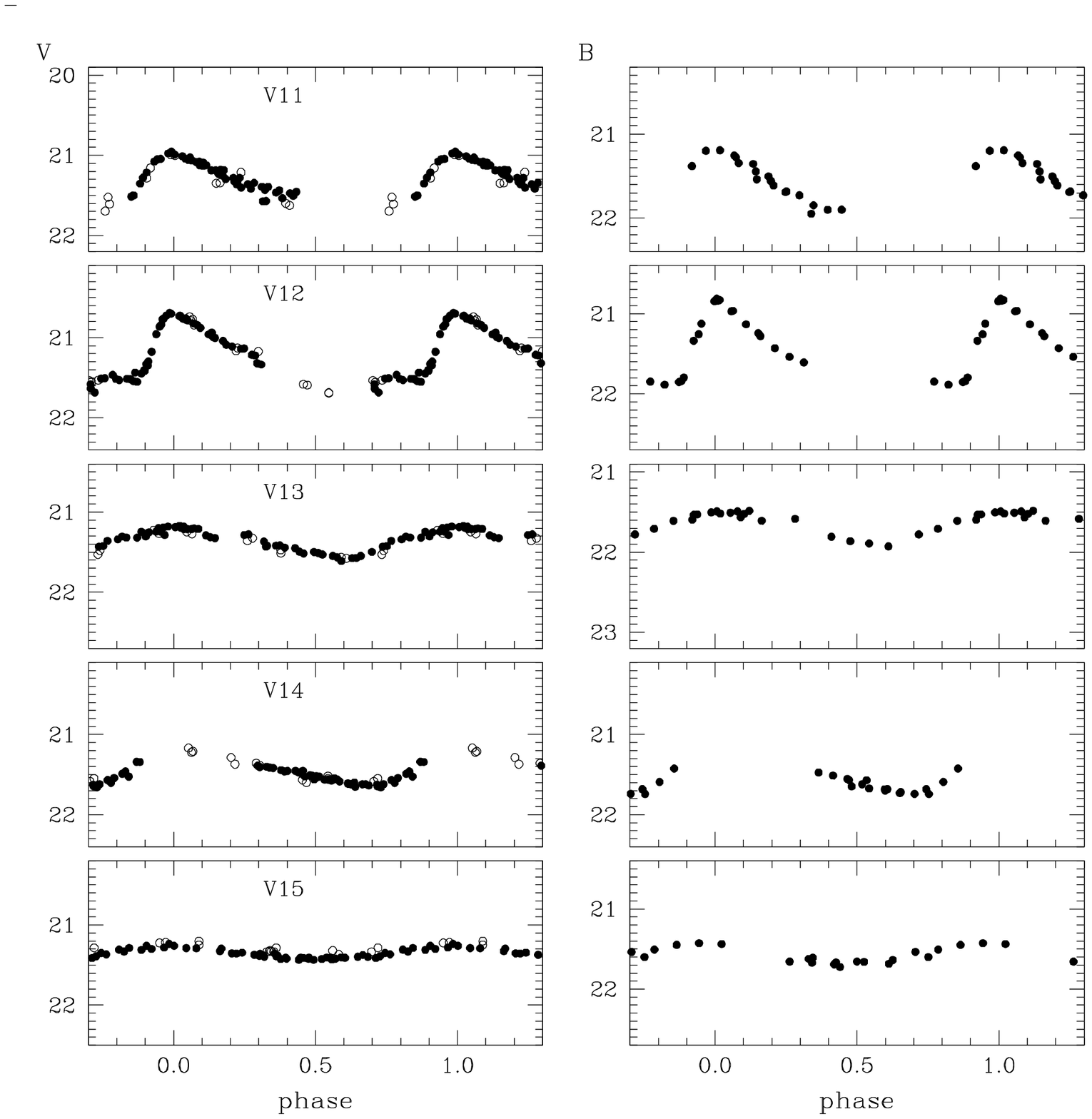}
\caption{Atlas of the $V$, $B$ light curves of the \rrl\ stars  in For~5
from the combined Magellan (filled circles) and HST (open circles) datasets.
Data for star V15 were folded adopting P=0.340 days (see Sect. 4.3).
}
\label{f:fig1a}
\end{figure*}

\begin{figure*} 
\figurenum{4}
\includegraphics[width=18cm,bb=40 159 580 703,clip]{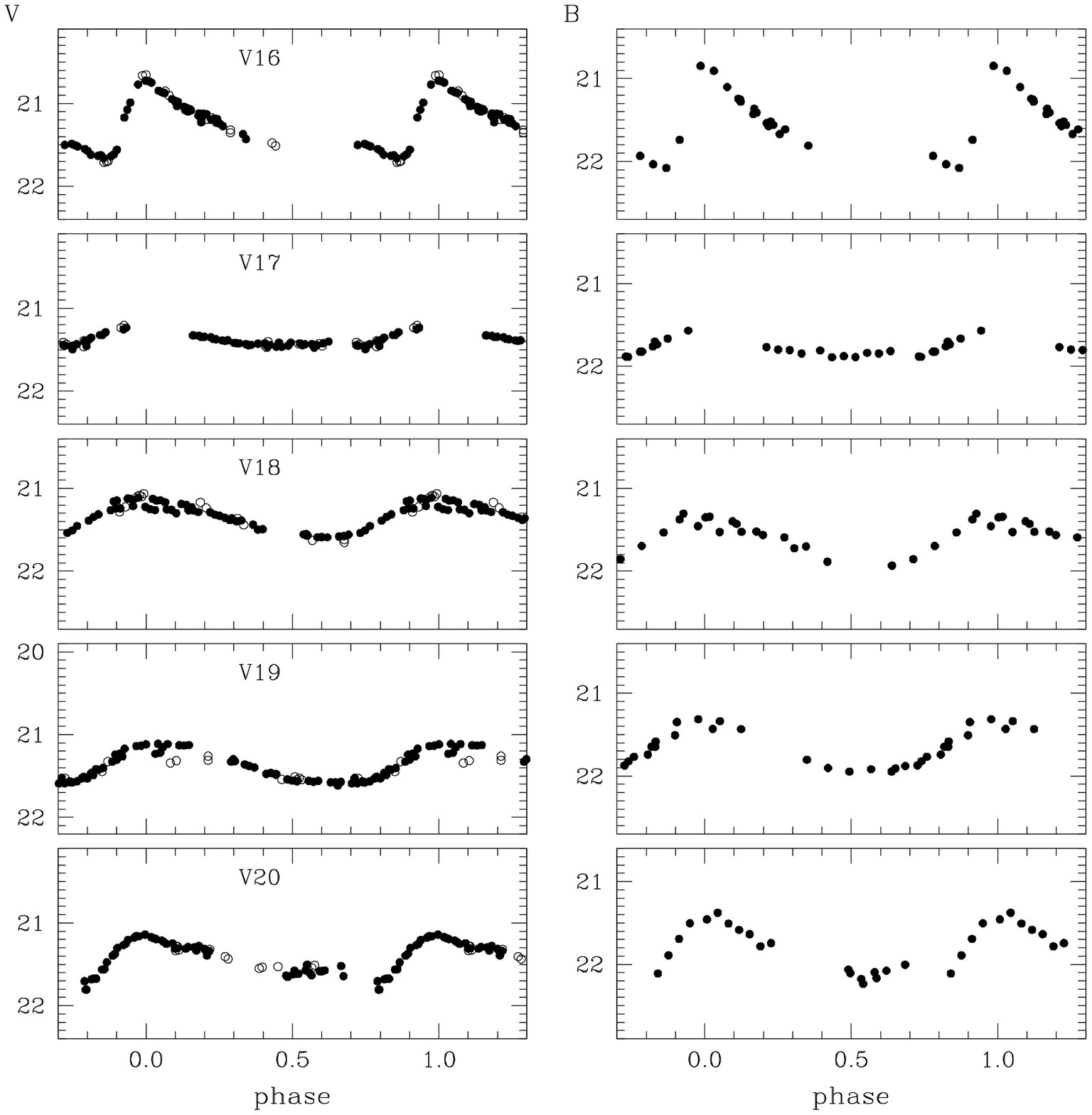}
\caption{continued}
\end{figure*}

\begin{figure*} 
\figurenum{4}
\includegraphics[width=18cm,bb=40 159 580 703,clip]{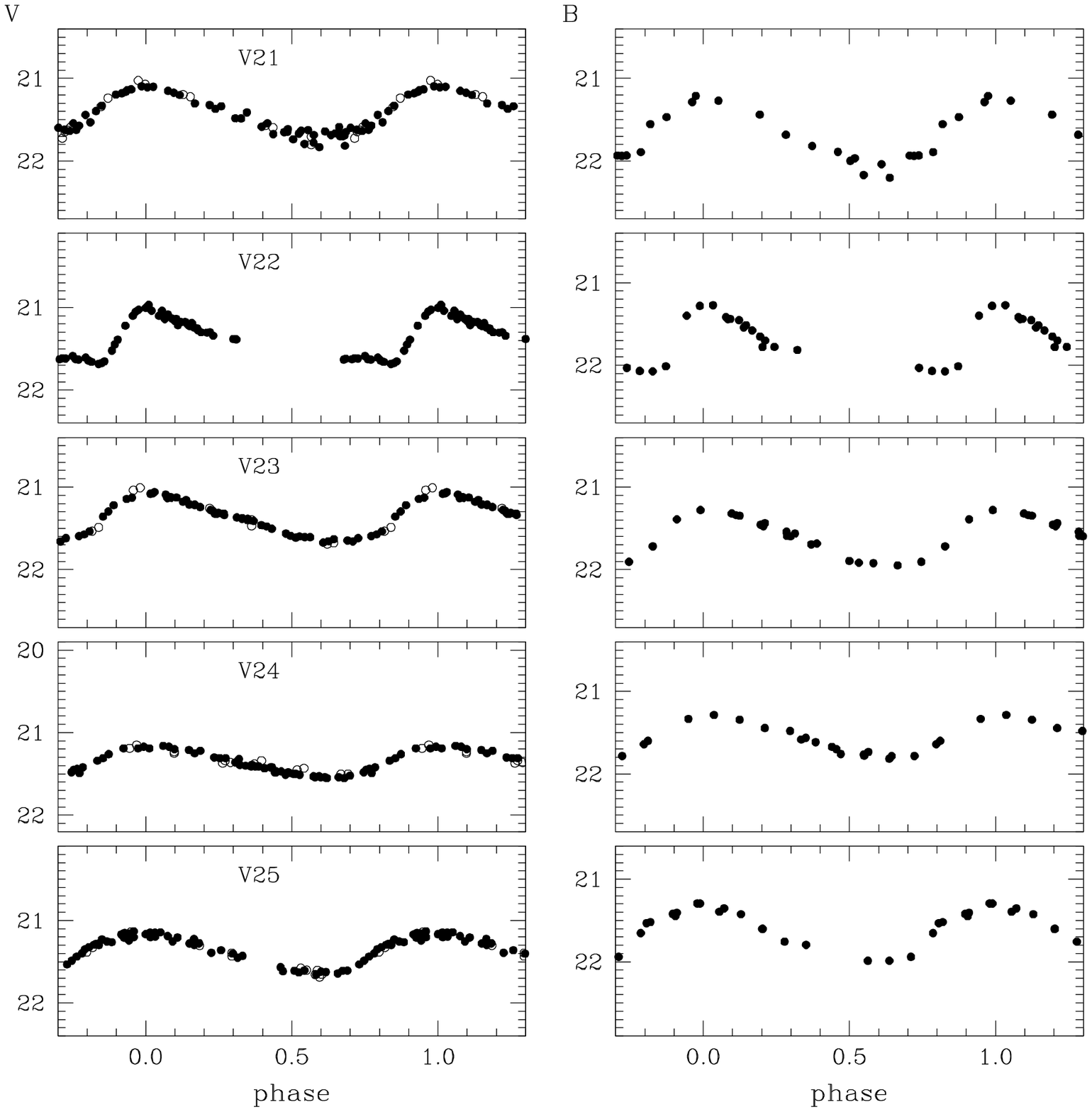}
\caption{continued}
\end{figure*}

\begin{figure*} 
\figurenum{4}
\includegraphics[width=18cm,bb=40 159 580 703,clip]{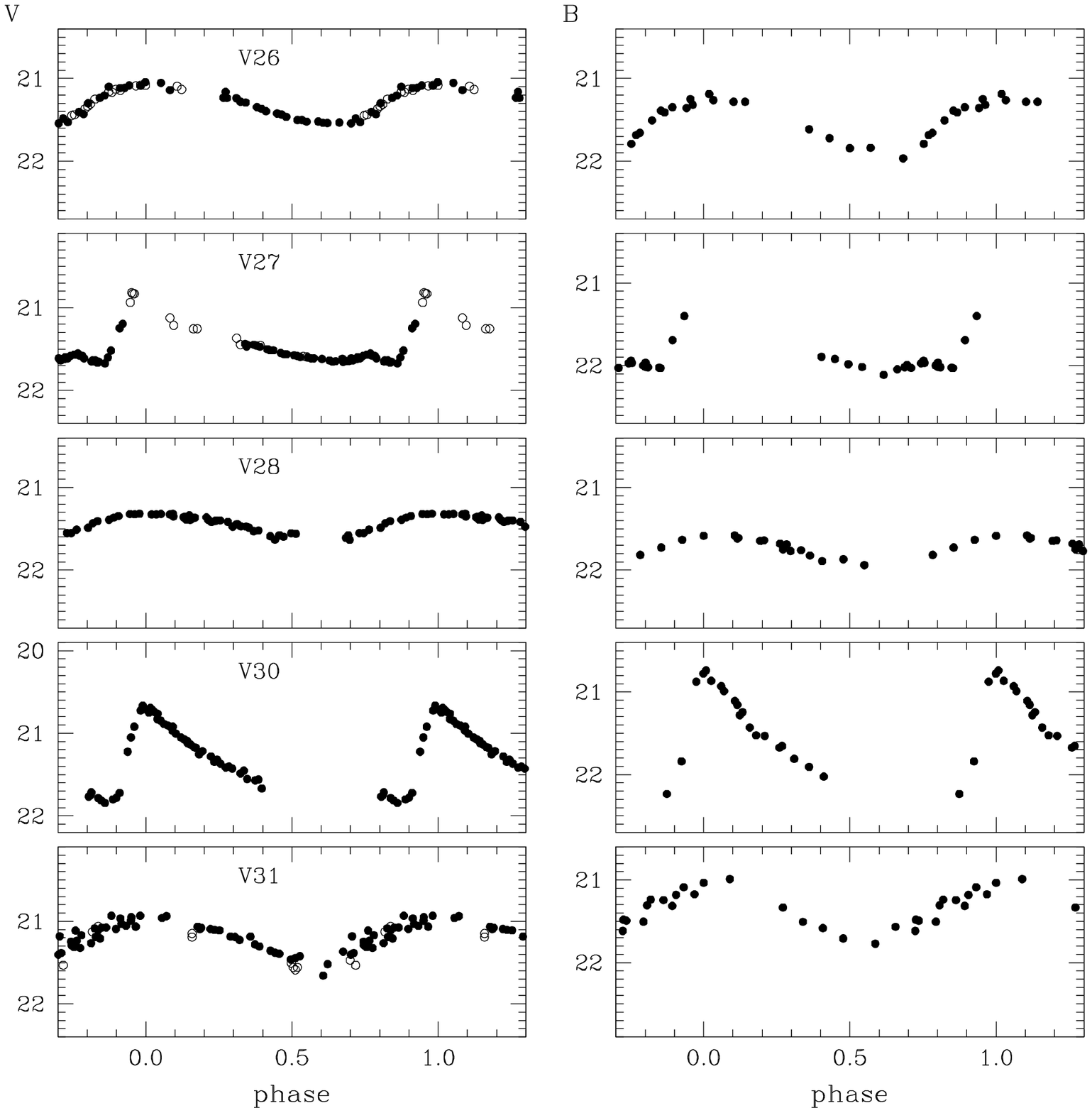}
\caption{continued}
\end{figure*}

\begin{figure*} 
\figurenum{5}
\includegraphics[width=16cm,bb=17 144 580 703,clip]{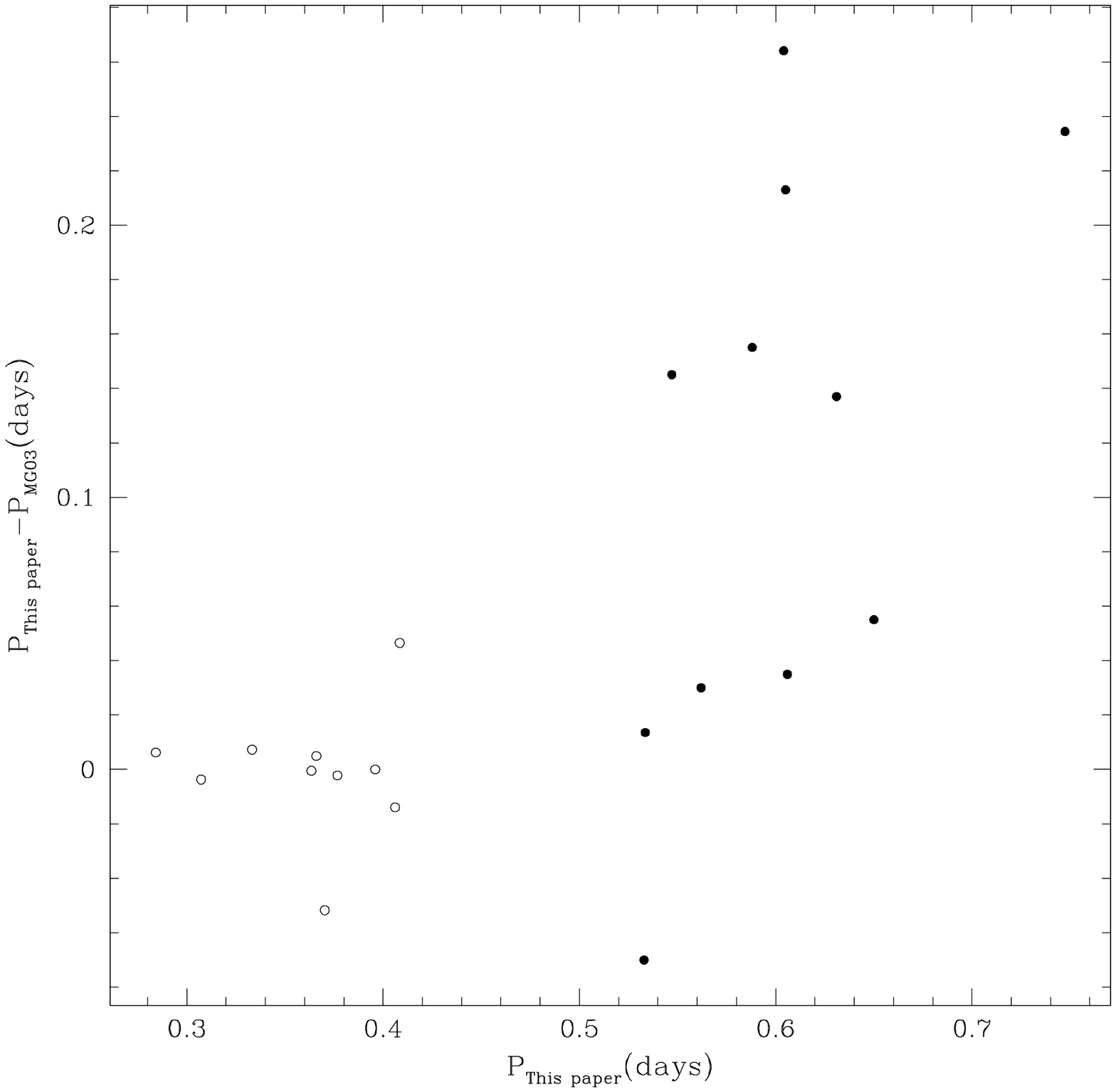}
\caption{Comparison between our period determinations and MG03 values for the stars
in common. Open and filled circles represent first-overtone and fundamental-mode pulsators, respectively,
according to our type classification.}
\end{figure*}

\begin{figure*}
\figurenum{6}
\includegraphics[width=14cm,bb=30 180 550 610,clip]{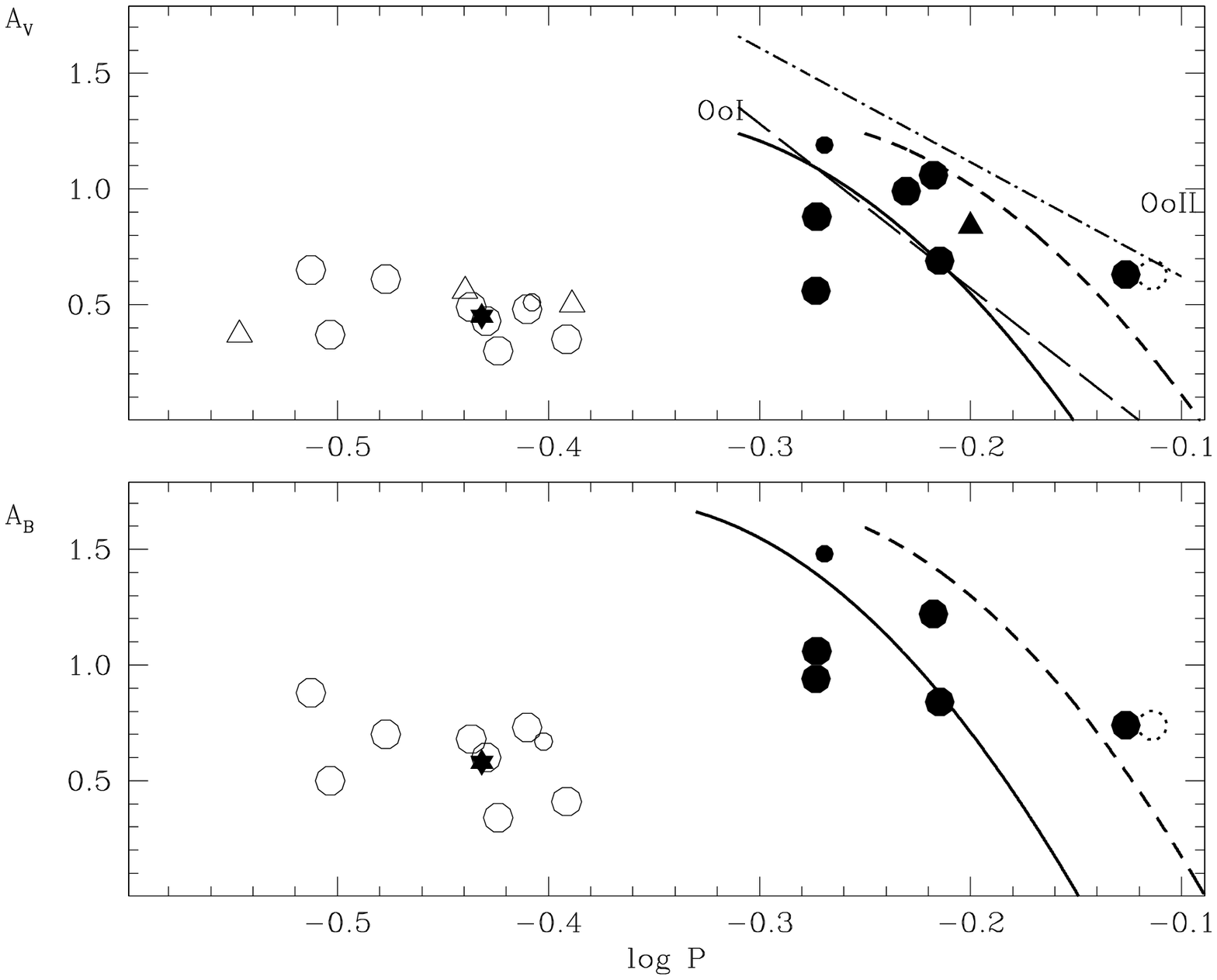}
\caption{$V$, $B$ period-amplitude diagrams of the For~5 
\rrl\ stars.
{\it Filled circles}: ab-type RR Lyrae stars; {\it open circles}: first-overtone RR Lyrae stars; 
{\it star}: candidate double-mode pulsator (RRd). Larger symbols indicate variable stars located at distances
less $45\arcsec$ from the cluster center. 
 {\it Filled} and {\it open triangles} are RRab and RRc variables within 8$^{\prime \prime}$ from the 
center of For~5 which have light curves from the HST observations only. The dotted open circle shows the position
of the RRab star V20 if the alternative longer period of 0.769393 days is adopted.
 The straight 
lines show the positions of the OoI and OoII Galactic GCs according to Clement \& Rowe (2000). 
Period-amplitude distributions of the
 bona fide regular ({\it solid curves}) and well-evolved ({\it dashed curves}) {\it ab-}type
  \rrl\ stars in M3 
from Cacciari, Corwin, \& Carney (2005) are also shown for comparison.
}
\end{figure*}

\end{document}